\documentclass[twocolumn]{aastex62}

\graphicspath{{./}{figures/}}

\received{March 8, 2019}
\revised{August 25, 2019}
\accepted{September 17, 2019}
\submitjournal{ApJ}

\shorttitle{Role gas fragmentation}
\shortauthors{Suazo et al.}

\begin{document}

\title{The Role of Gas Fragmentation During the Formation of Supermassive Black Holes}

\correspondingauthor{Mat\'ias Suazo}
\email{matias.suazo@ug.uchile.cl}

\author[0000-0002-0786-7307]{Mat\'ias Suazo}
\affil{Departamento de Astronom\'ia, Universidad de Chile, Casilla 36-D, Santiago, Chile.}

\author{Joaqu\'in Prieto}
\affiliation{Departamento de Astronom\'ia, Universidad de Chile, Casilla 36-D, Santiago, Chile.}

\author{Andr\'es Escala}
\affiliation{Departamento de Astronom\'ia, Universidad de Chile, Casilla 36-D, Santiago, Chile.}

\author{Dominik R.G. Schleicher}
\affiliation{Departamento de Astronom\'ia, Facultad Ciencias F\'isicas y Matem\'aticas, Universidad de Concepci\'on, Av. Esteban Iturra s/n Barrio Universitario, Casilla 160-C, Concepci\'on, Chile}



\begin{abstract}


We perform cosmological hydrodynamic simulations to study the effect of gas fragmentation on the formation of supermassive black hole seeds in the context of Direct Collapse. Our setup considers different UV background intensities, host halo spins, and halo merger histories. We observe that our low-spin halos are consistent with the Direct Collapse model when they are irradiated by a UV background of $J_{21} = 10000$. In these cases, a single massive object $\sim10^5$ $M_{\odot}$ is formed in the center of the halo. On the other hand, in our simulations irradiated by a UV background of $J_{21} = 10$, we see fragmentation and the formation of various less massive seeds. These fragments have masses of $10^3$ - $10^4$ $M_{\odot}$. These values are still significant if we consider the potential mergers between them and the fact that these minor objects are formed earlier in cosmic time compared to the massive single seeds. Moreover, in one of our simulations, we observe gas fragmentation even in the presence of a strong UV intensity. This structure arises in a dark matter halo that forms after various merger episodes, becoming the structure with the highest spin value. The final black hole seed mass is $\sim 10^5$ $M_{\odot}$ for this run. From these results, we conclude that fragmentation produces less massive objects; however, they are still prone to merge. In simulations that form many fragments, they all approach the most massive one as the simulations evolve. We see no uniqueness in the strength of the UV intensity value required to form a DCBH since it depends on other factors like the system dynamics in our cases.

\end{abstract}



\section{Introduction} \label{sec:intro}

Quasars are one of the most intriguing astrophysical objects. They are extremely luminous active galactic nuclei powered by supermassive black holes (SMBHs) in their centers. Observations of $z > 6$ quasars \citep{fan2006,mortlock2011,wu2015,bañados,dominik_book} show that these objects already hosted SMBHs with masses up to $10^9$ $M_{\odot}$ when the universe was $\sim$$0.7$ Gyrs old. The formation of such massive objects is still an open question. Various scenarios have been proposed to explain the early assembling of these structures, including primordial stellar remnants, stellar dynamical processes, and the direct collapse of protogalactic gas due to dynamic instabilities. For further reading, see \citet{volonteri2010}, \citet{volonteri2012}, and \citet{latif_rev}.

The standard model of structure formation states that there is a population of stars made of primordial metal-free gas, these are the Population III of stars (\citealt{bromm1999,bromm2002,abel2000,abel2002,yoshida2006}). Recent observations of the 21 cm absorption line due to its interaction with Lyman-$\alpha$ photons reveal that the first stars were formed at $z$ $\sim$15 - 20 \citep{bowman}. Population III stars are expected to be formed in minihalos of 10$^5$ - 10$^6$ $M_{\odot}$. In these halos, molecular hydrogen acts as an efficient coolant, in which star formation takes place. Because these stars are formed out of metal-free gas, they are expected to be more massive than observed local stars. Three-dimensional cosmological simulations show that Population III stars can reach masses of a few hundred $M_{\odot}$ \citep{hirano,susa}. Depending on the stellar mass, their fate is different, but most of them end up as stellar black hole remnants \citep{heger}. It is known that radiative feedback effects halt the growth of these structures \citep{johnson,alvarez}. Stellar black holes would require several super Eddington accretion episodes to reach a billion solar masses by redshift $\sim 6-7$ \citep{pacucci2015,volonteri2015}.

The formation of a very massive star (VMS) in dense stellar clusters is also a candidate for a supermassive black hole seed. If the center of a cluster undergoes core collapse, a single massive compact object may form. The core collapse timescale has to be shorter than the typical timescale for the massive stars to reach the main
sequence, to avoid supernovae feedback. \citep{devecchi2009,lupi_2014,sakurai_2017,stone2017,boekholt__2018,reinoso_2018}. 

The formation of a massive seed has also been considered in the Direct Collapse Black Hole (DCBH) model. It consists of the immediate collapse of a gas cloud into a massive seed of 10$^4$ - 10$^6$ $M_\odot$. In this model, the seed would ultimately collapse into a black hole more massive than the ones produced by other models. The seeds would accrete at rates $\geq 0.1$ $M_\odot$yr$^{-1}$ \citep{begelman10,dominik13,hosokawa2015}. However, this scenario demands strict conditions. A very massive halo is indeed necessary to have the gas reservoir to form a massive seed and also to ensure a high gas temperature ($\sim$$8000$ K). Such high temperatures are required to allow atomic cooling to operate. In comparison to molecular and metal cooling, atomic cooling is characterized to act more smoothly in the temperature ranges involving DCBH conditions, permitting an isothermal collapse that does not produce fragments \citep{latif2013a,latif2014}.
H$_2$ inhibition is fundamental in the direct collapse model. This molecule can cool the system down to hundreds of Kelvins producing fragments and the formation of minor objects, preventing a single massive body to be assembled. The main mechanism to form molecular hydrogen is shown in Equation~\ref{eq:formdos} which requires the previous formation of H$^{-}$ that is shown in Equation~\ref{eq:formuno}.

\begin{equation}
\text{H} + \text{H}^{-} \rightarrow \text{H}_{2} + \text{e}^{-},
\label{eq:formdos}
\end{equation}

\begin{equation}
\text{H} + \text{e}^{-} \rightarrow \text{H}^{-} + \gamma .
\label{eq:formuno}
\end{equation}


One way to prevent the existence of molecular hydrogen is through the direct exposition to photons with energies between 11.2 and 13.6 eV in the Lyman--Werner (LW) band. The chemical reaction is illustrated in Equation~\ref{eq:lw}, where H$_2^{*}$ represents an excited state for molecular hydrogen. Another way to avoid molecular hydrogen formation is preventing the formation of H$^{-}$, recognized to catalyze H$_2$ formation. That can be obtained through the interaction of low energy photons (above 0.76 eV) that photodetach the H$^{-}$, which impedes the major H$_2$ formation channel in the early Universe (Equation~\ref{eq:detach}).

\begin{equation}
\text{H}_2 + \gamma _{LW} \rightarrow \text{H}_2 ^{\star} \rightarrow \text{H} + \text{H},
\label{eq:lw}
\end{equation}


\begin{equation}
\text{H}^{-} + \gamma _{0.76} \rightarrow \text{H} + e^{-}.
\label{eq:detach}
\end{equation}

\vspace{3mm}


The conditions to prevent H$_2$ formation can be fulfilled if there is a star-forming galaxy irradiating the aforementioned photons next to a pristine halo. A galaxy with a spectrum of $T_{\text{eff}} = 10^5$ K is more efficient at dissociating H$_2$ directly (Equation~\ref{eq:lw}) while a spectrum of $T_{\text{eff}} = 10^4$ K is more efficient indirectly by destroying H$^{-}$ (Equation~\ref{eq:detach}). In principle, molecular hydrogen inhibition requires a critical value of the UV flux ($J_{21}^{\text{crit}}$) that depends on the spectrum. UV intensities are typically specified in units of J$_{21} = 10^{-21} \text{erg } \text{s }^{-1} \text{Hz }^{-1}\text{cm }^{-2}\text{str }^{-1}$. \citet{latif2014a} quantifies the strength of J$_{21}^{\text{crit}}$ by performing 3D cosmological simulations, finding this value to range 400 - 700. Recent estimates for 3D cosmological simulations including realistic Population II radiation fields and X-rays find that J$_{21}^{\text{crit}}$ varies between 20000 and 50000 \citep{latif2015c}. In addition, \citet{glover2015} argues that the chemical network used in the simulations impacts directly on the value of J$_{21}^{\text{crit}}$, so he constructed a reduced network of 26 chemical reactions using a reaction-based reduction technique by performing one-zone models. Under such results, there is no a clear agreement on J$_{21}^{\text{crit}}$, since this value depends on local conditions, such as density, temperature, chemical composition, and its dynamics. In extreme cases, i.e. when cosmological halos that collapse very rapidly, a UV background may not even be necessary, as the compressional heating of the gas provides an alternative channel for the destruction of H$_2$ by collisional dissociation \citep{wise_2019}. Here we rather address the possibility that H$_2$ cooling may occur and that massive objects could nevertheless form.

Though the DCBH model can explain the formation of massive seeds, there are still some obstacles that the gas needs to overcome. One obstacle arises when gas is photoevaporated, in which case a dense core is prevented from forming. Ionization also impedes the formation of a central structure, since it leads to an increase in the electron fraction, which is one of H$_2$ catalysts \citep{johnson2014}. Another obstacle corresponds to the tidal interaction that the neighborhood may be exerting on the pristine halo, that could disrupt it, avoiding the formation of a massive central object \citep{chon}.

The goal of this paper is to analyze the effect of fragmentation on the formation and growth of SMBH seeds, considering the formation of minor objects and their merger events, in more flexible conditions that are more likely to happen.

\section{Methods}

We perform a set of nine simulations in box sizes of 1 cMpc $h^{-1}$ using the adaptive mesh refinement hydrodynamical cosmological code RAMSES \citep{ramses}. Initial conditions are created using the MUSIC code \citep{music} that generates random Gaussian initial perturbations and allows the usage of the  `zoom-in' technique, in which a small region encloses the formation history of the object of interest, so it is studied with a much higher resolution. By doing this, we obtain a maximum resolution of $\sim$0.01 pc in proper units. We use the cosmology given by the \citet{planck} that sets $\Omega_b = 0.3089$, $\Omega_{\Lambda} = 0.6911$, $\sigma_{8} = 0.8159$, $n_s$ = 0.9967, $H_{0} = 67.74$ km s$^{-1}$ Mpc$^{-1}$. The Jeans length is resolved according to the Truelove criterion \citep{truelove} to avoid numerical fragmentation. We use 32 cells per Jeans length as the minimum reasonable value suggested by \citet{latif2013} to resolve turbulences. 

\subsection{Chemical Model}

We solve the thermal and chemical evolution of the gas employing the publicly available package KROME \citep{krome}. We use primordial chemistry made up of the following species: H, H$^{+}$, H$^{-}$, He, He$^{+}$, He$^{++}$, $e^{-}$, H$_2$, H$_2^{+}$. We also included one extra run with additional deuteride species: D, D$^{+}$, HD, HD$^{+}$, D$^{-}$. We assume a uniform isotropic UV background of various intensities that would come from a star-forming neighborhood. The spectra of those regions are a blackbody with $T_{\text{eff}} = 2 \times 10^{4}$ K, which is in the range that mimics realistic spectra \citep{sugimura,latif2015c}. All chemical reactions with their reaction rates are listed in Appendix~\ref{sec:chem}, which includes the photodetachment of H$^{-}$, photodissociation of H$_2$ and H$_2^{+}$, collisional dissociation, collisional induced emission, chemical cooling/heating from three-body reactions, cooling by collisional excitation, collisional ionization, radiative recombination, and bremsstrahlung radiation. We also add the H$_2$ self-shielding fitting function by \citet{wg11}, where we compute the molecular hydrogen column density as it is shown in Equation~\ref{ec:nh2}, where n$_{\text{H}_2}$ is the H$_2$ number density. The $max$ function is employed to avoid numerical artifacts.

\begin{equation}
    \text{N}_{\text{H}_{2}} = 2 \times 1.87 \cdot 10^{21} (max(n_{\text{H}_2},10^{-40})\cdot 10^{-3})^{2/3}
    \label{ec:nh2}
\end{equation}

\subsection{Sink Particles}
\label{sec:sink}

The gravitational collapse of dense regions is a recurring phenomenon in astrophysical simulations, therefore, a huge dynamical range is required. However, computational resources are not always sufficient to resolve the small scales.


Sink particles are particles that approximate the scales that are not resolved by the collapse of a region into a single point. This point is disconnected from the hydrodynamics, it interacts with the remaining gas through gravity and accretion only and with other sink particles that are formed. Different schemes to create sink particles have been formulated to make them represent more realistic physical phenomena. \citet{bleuler} present the algorithm implemented in the RAMSES code to create sink particles, which is based on a clump finder. Unlike other techniques to form sink particles, this scheme requires a region to have cells denser than a given threshold and also have a certain `prominence,' which is the ratio between the peak density and the maximum saddle density to this peak. By fulfilling the aforementioned requirements and other three physical checks, a sink particle forms. 

We use the checks presented in \citet{latifvolonteri} to decide if a clump will become a sink particle. One of the conditions is the selection of regions whose peaks are at the highest refinement level, and the other condition requires having an excess in mass relative to the Jeans mass for the candidate region to become a sink. A third check consists of the inability to create a sink particle if its formation would occur inside the accretion radius of a sink particle previously formed sink particle. Once all checks are fulfilled in a `prominent' region, the sink particle is formed and its initial mass is calculated such that the cell from which the particle is formed remains Jeans stable after the subtraction of the sink mass. Sink particles emulate accreting SMBH seeds in this work.

Sink particle trajectories are integrated through the direct force summation scheme presented in \citet{bleuler}. Their acceleration due to the interaction with gas and other sink particles is computed by looping over all pairwise combinations and calculating their mutual attraction. In both cases our particles are softened using a Plummer profile, being the softening radius two times the smallest cell in the simulation.

Sink particles are allowed to merge if they are at a distance closer than four times the smallest cell in the simulation; this distance is known as the accretion radius. This criterion has been extensively used in simulations with sink particle schemes (e.g. \citet{dubois_2010}). The accretion in these sink particles follows the `flux accretion' scheme presented in \citet{bleuler}. It basically computes Equation~\ref{ec:flux}, where $\Omega_{\text{acc}}$ is a sphere defined by the accretion radius, $\rho$ is gas density, $\vec{v}$ is gas velocity, and $\vec{v}_{\text{sink}}$ is the sink particle velocity.

We perform several checks to ensure that the motion of the sink particles and their interaction is physical, These checks are presented in Appendix~\ref{sec:sink_a}.


\begin{equation}
    \dot{M}_{\text{flux}} = - \int_{\Omega_{\text{acc}}} \nabla \rho (\vec{v} - \vec{v}_{\text{sink}}) dV
    \label{ec:flux}
\end{equation}


\section{Results}
\label{sec:results}

\begin{figure}
\includegraphics[width=\columnwidth]{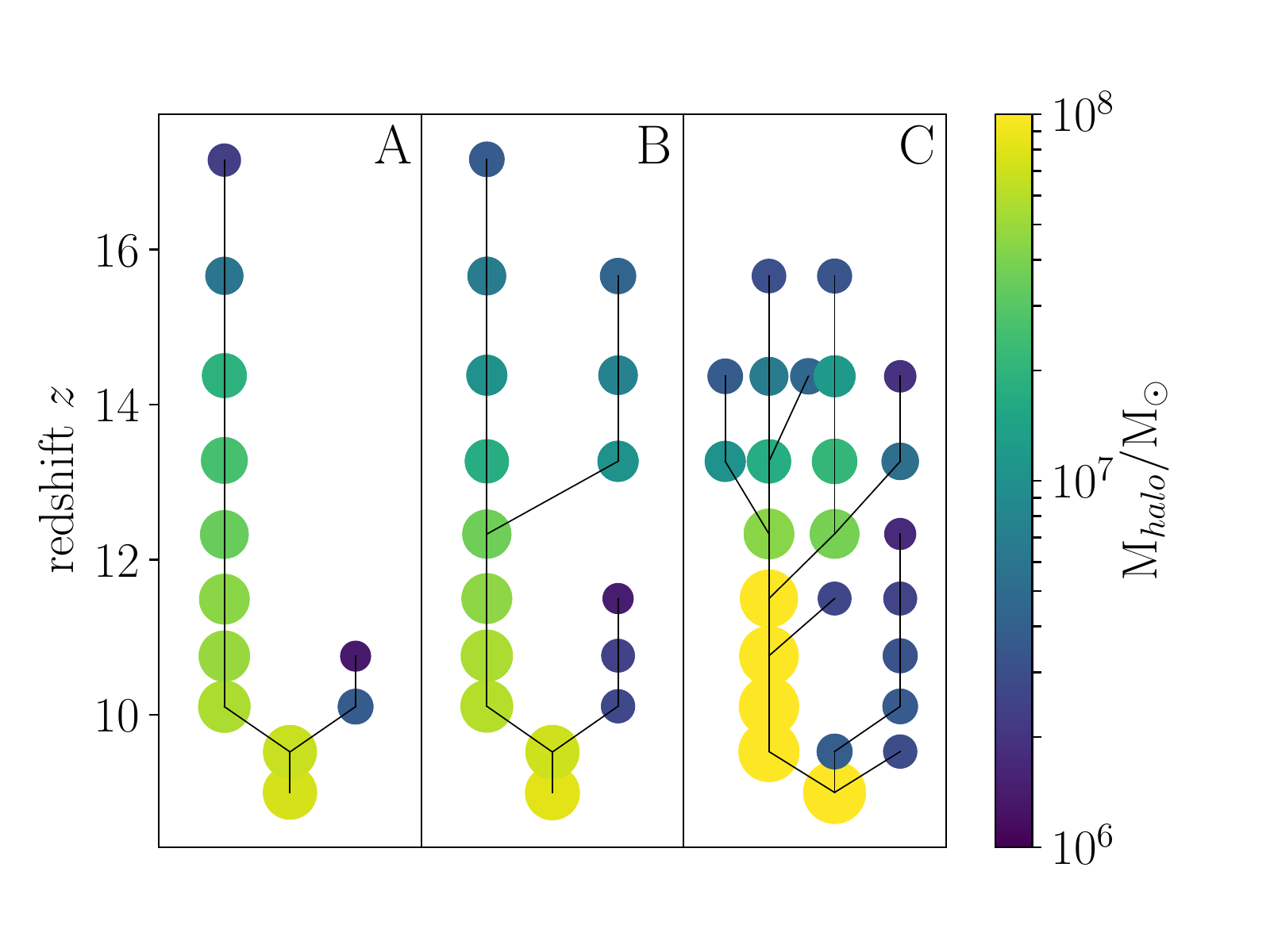}
\caption{Merger trees for selected halos, where evolution goes downwards. Halos A, B, and C are shown from left to right. Masses are represented with colors and circle sizes. The y-axis represents redshift while the x-axis is dummy.}
\label{fig:merger_tree}
\end{figure}

We perform 3D simulations to study the effect of gas fragmentation in the formation of primordial SMBHs in three different halos. We start from only dark matter (DM) low-resolution simulations initialized at $z = 99$ in 1 cMpc $h^{-1}$ box sizes to re-simulate some of them. We make our selection based on three criteria: (i) halo mass, (ii) halo spin parameter, and (iii) merger history. Halo masses are required to be higher than $\sim 5 \times 10^{7}$ $M_{\odot}$ to guarantee the virial temperature to be in the atomic cooling regime ($>$ 8000 K, e.g. \citeauthor{barkana}, \citeyear{barkana}), in which gas is mainly cooled down due to electronic transitions in the hydrogen atom. Usually, $J_{21}$ is used as a unique parameter to determine the feasibility of a DCBH scenario. However, it is too simplistic, since the dynamics affect the chemistry and the conditions to allow a scenario. This is why we also make our choice based on selecting diverse halo spin parameters and by checking the merger status using merger trees. Massive isolated halos are rare to find in 1 cMpc $h^{-1}$ box sizes, so we run 50 simulations and pick three of them according to their properties at $z \sim 9$. We identify DM halos using the HOP clump finder \citep{hop}. The chosen halos are denominated A, B, and C and their properties, such as the Bullock spin parameter \citep{bullock}, virial mass, and the number of DM particles, are shown in Table~\ref{tab:halo_prop}. In Figure~\ref{fig:merger_tree}, we show the merger trees for all selected halos. The number of halo mergers is one for A and two for B. The mass ratio for the only merger in A is 0.067 with the lowest spin parameter value, while for B mass ratios are 0.60 and 0.04 (top and bottom, respectively, according to the figure). Halo C turns to be the one formed out of a high number of mergers. This phenomenon seems to be necessary to ensure a high spin parameter \citep{peirani}. Equation~\ref{ec:bullock} represents the Bullock spin parameter formula, where $|\vec{J}|$ is the magnitude of the total angular momentum of the halo, $M$ is the virial mass of the halo, and $V$ is the virial velocity defined as $V = \sqrt{GM/R}$, where $R$ is the virial radius of the halo. \citet{bullock} show that the spin parameter follows a log-normal distribution as it is shown in Equation~\ref{ec:log}, with $\mu$ = 0.035 and $\sigma$ = 0.5.
\begin{equation}
    \lambda \equiv \frac{|\vec{J}|}{2MRV}
    \label{ec:bullock}
\end{equation}

\begin{equation}
p(x) = \frac{A}{x\sqrt{2\pi}\sigma}\exp \left(-\frac{\ln^2 (x/\mu)}{2\sigma ^2}\right)
\label{ec:log}
\end{equation}

\begin{figure}
\includegraphics[width=\columnwidth]{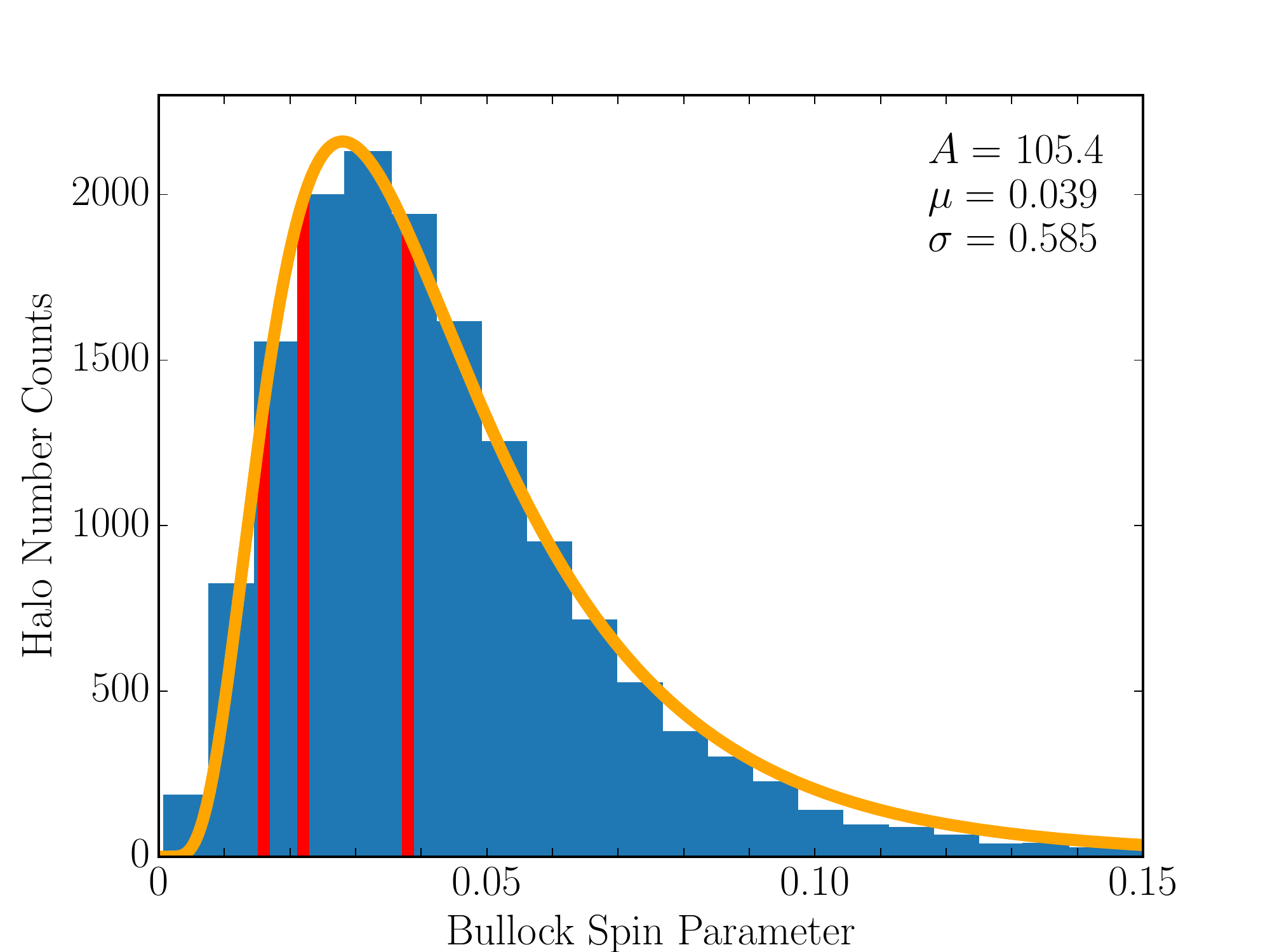}
\caption{Spin parameter distribution for all halos found in our 50 DM-only low-resolution simulations. The orange line represents a log-normal fitting function, while the red lines represent the spin parameters for the selected halos. A, $\mu$, and $\sigma$ represent the best fitting parameters according to Equation~\ref{ec:log}.}
\label{fig:spin}
\end{figure}

In Figure~\ref{fig:spin}, we show the spin distribution for all halos found in our DM-only low-resolution simulations, where the orange line represents the best log-normal fitting function and the red lines represent the spin parameters for Halos A, B, and C from left to right, respectively. We obtain values of $A = 60.34$, $\mu = 0.039$, and $\sigma = 0.582$ from the fitting, implying a mean and a standard deviation of 0.046 and 0.029, respectively. Spin parameters are chosen to lie at 1.02, 0.82, and 0.27 standard deviations from the mean value in the function. All chosen values are below the mean and represent the most relevant cases for the DCBH scenario.

Once the halos are selected, they are re-simulated including gas physics, also making use of the KROME package and the zoom-in technique. We also add our modified algorithm to form sink particles (see Section~\ref{sec:sink}). The zoom-in is made by enclosing a spherical region of a size equal to the virial radius of the respective halo. Such a choice does not affect our final results since all structures are formed at scales orders of magnitude below the virial radius. In all cases, the virial radius has a value of $\sim $2 kpc. For all re-simulations, we add a uniform UV background with intensity values of $J_{21} = 10000$ and $J_{21} = 10$. In Appendix~\ref{sec:j21} we show other cases, which are quite similar to the ones shown in the next sections. 

In Table~\ref{tab:halo_sink} we summarize some of the sink features for our results; the redshift at the moment at which the first sink particle formed and its corresponding cosmic age, the total number of sink particles formed during the run, the number of sink particle mergers, the initial first sink particle mass, the final first sink particle mass, and the average accretion rate for the first sink particle considering the first 300 kyr since its creation in the corresponding run, and also the rest of the run. The simulations A, B, and C in the presence of a $J_{21}$ = 10000 are run for 360, 520, and 620 kyr, respectively, after the formation of the first sink particle. The simulations A, B, and C in the presence of a $J_{21}$ = 10 are run for 3032, 3950, and 3062 kyr, respectively, after the formation of the first sink particle.

From Table~\ref{tab:halo_sink}, we observe that sink particles are formed earlier in cosmic time when we decrease the UV intensity. This can be understood from the fact that $M_{\text{jeans}}$ $\propto$ $T^{3/2}$. At low UV intensities, the gas is cooled down, so the thermal jeans mass decreases and regions collapse sooner compared to high UV intensities, in which the gas remains hot and allows the assembly of baryons for a longer period. This delay in gas assembly for high UV intensities might have an important consequence for seed formation.

\begin{table}
\centering
\caption{Virial Mass, Spin Parameter, and Number of Particles in the Low-Resolution DM-only Simulated Halos.}
\label{tab:halo_prop}
\begin{tabular}{| c | c | c | c |}
\hline
Halo & Virial Mass [M$_{\odot}$] & Spin Parameter & n$_{\text{part}}$\\
\hline
A & $9.68 \times 10^{7} $ & 0.016 & 1612 \\
B & $4.90 \times 10^{7} $ & 0.022 & 821 \\
C & $2.68 \times 10^{8} $ & 0.038 & 4471 \\
\hline
\end{tabular}
\end{table}

\begin{table*}
\centering
\caption{Sink features for the simulations: redshift and cosmic age for the first sink formed, number of sink particles formed during the run, number of mergers during the run, initial mass for the first sink particle created, average accretion rate for the first sink particle formed in the run considering the first 300 kyr, and accretion run considering the whole run.}
\label{tab:halo_sink}
\begin{tabular}{| c | c | c | c | c | c | c | c | c | c |}
\hline
Halo & J$_{21}$ & $z$ & Cosmic & N$^{\circ}$ sinks & Sink & Initial sink & Final sink & Average accretion & Average accretion \\
& & & age (Myr) & formed & mergers & mass ($M_{\odot}$) & mass ($M_{\odot}$) & 300 kyrs ($M_{\odot}$yr$^{-1}
$) & rate ($M_{\odot}$yr$^{-1}$) \\
\hline
A & 10000 & 11.28 & 400.1 & 1 & - & 337 & 1.0 $\times 10^5$ & 0.331 & 0.279 \\ 
  & 10 & 13.11 & 324.4 & 4 & 1 & 13 & 1.7 $\times 10^4$ & 0.009 & 0.005 \\ 
B & 10000 & 10.08 & 466.9 & 1 & - & 518 & 2.2 $\times 10^5$ & 0.574 & 0.374 \\ 
  & 10 & 11.03 & 412.5 &  13 & 2 & 18 & 9.7 $\times 10^3$ & 0.008 & 0.002 \\ 
C  & 10000 & 10.80 & 424.9 & 5 & 1 & 198 & 1.2 $\times 10^5$ & 0.080 & 0.128\\
  & 10 & 12.37 & 351.9 & 6 & - & 368 & 1.6 $\times 10^4$ & 0.007 & 0.005 \\
\hline
\end{tabular}
\end{table*}

\subsection{$J_{21} = 10000$}
\label{sec:j21_10000}

Results vary depending on every halo and UV intensity. Halos with a high UV intensity are expected to behave in similar ways, as H$_2$ is expected to be inhibited, and a DCBH is predicted to form \citep{shang2010,latif2014}.

\subsubsection{Dynamics}
\label{sec:j21_10000_dynamics}

We perform simulations for each halo using a uniform UV background of $J_{21} = 10000$. We follow the evolution during roughly 350 kyr after the first sink particle forms, which corresponds to a couple of growth times $t_{\text{growth}} = M_{\text{sink}}/\dot{M}_{\text{sink}}$ for these runs. This characteristic time $t_{\text{growth}}$ varies for all runs and sink particles, but during the first 350 kyr, this value is kept roughly constant in the range $150 - 200$ kyr. This value increases after this period, as a consequence of the increase in the sink mass and the decrease in the accretion rate as we show in Section~\ref{sec:rad_prof_10000}.

\begin{figure*}
\includegraphics[width = 2.1\columnwidth]{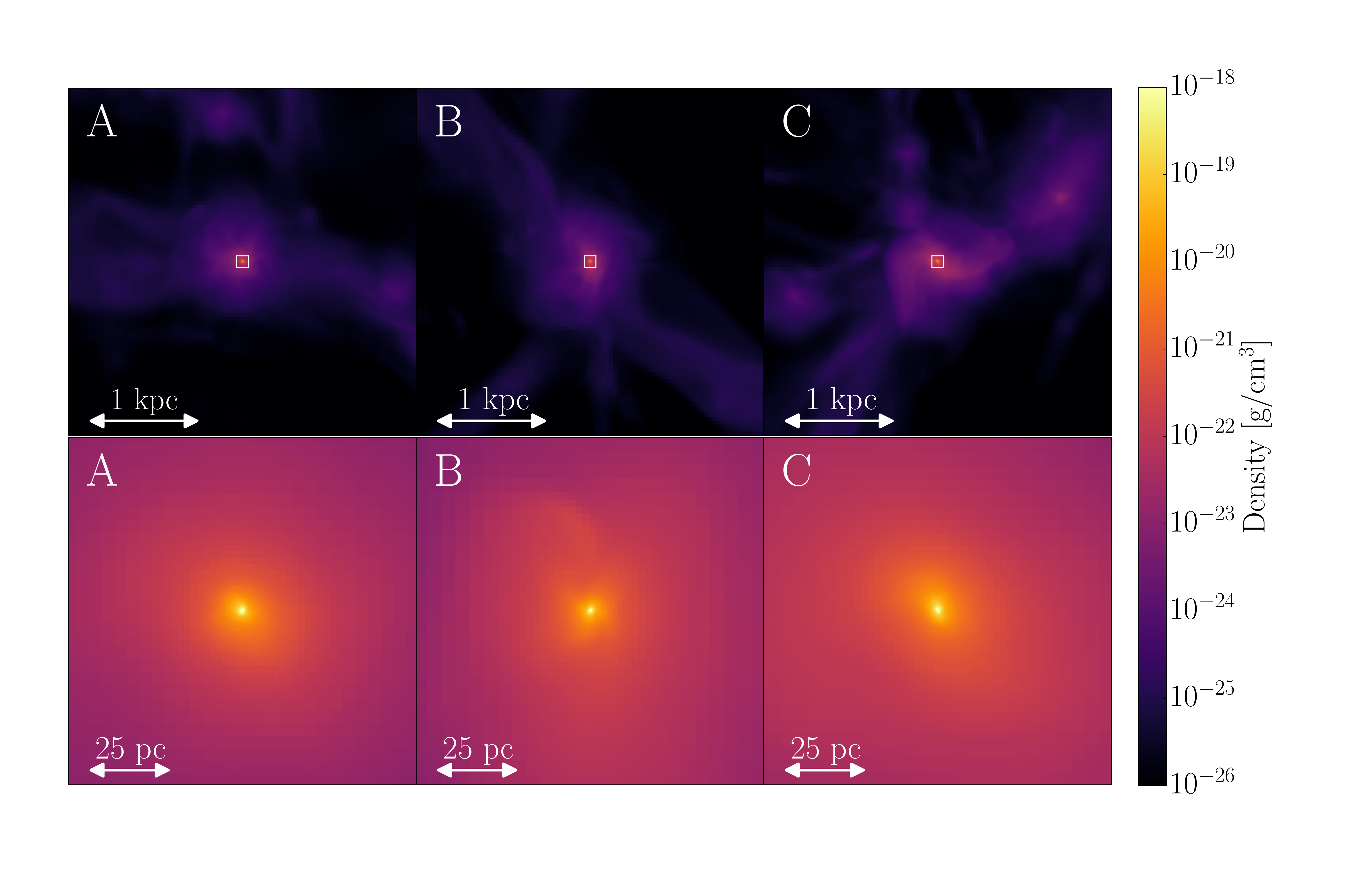}
\caption{Density projection for all re-simulated halos irradiated by a UV background of $J_{21} = 10000$ at the time when the runs finish. 
Halos A, B, and C are shown from left to right, respectively. At the top, projections correspond to 3 kpc, where white squares correspond to 100 pc regions, which are zoomed at the bottom.}
\label{fig:j21_10000_v_100}
\end{figure*}

Figure~\ref{fig:j21_10000_v_100} represents projections of all halos at a scale of 3 kpc (top) and a scale of 100 pc (bottom) at the end of each respective run. The projections for Halos A and B are centered on the only sink particle formed, while for C it is in the first sink particle created. We observe the formation of a dense central spherical structure at both large and small scales. At large scales, we observe some filaments feeding with gas every halo. This picture does not change with time. At the center of these structures, sink particles are created (not shown in the plots) and no significant dynamical evolution nor fragmentation is observed during the run at these scales.

\begin{figure*}
\includegraphics[width = 2.1\columnwidth]{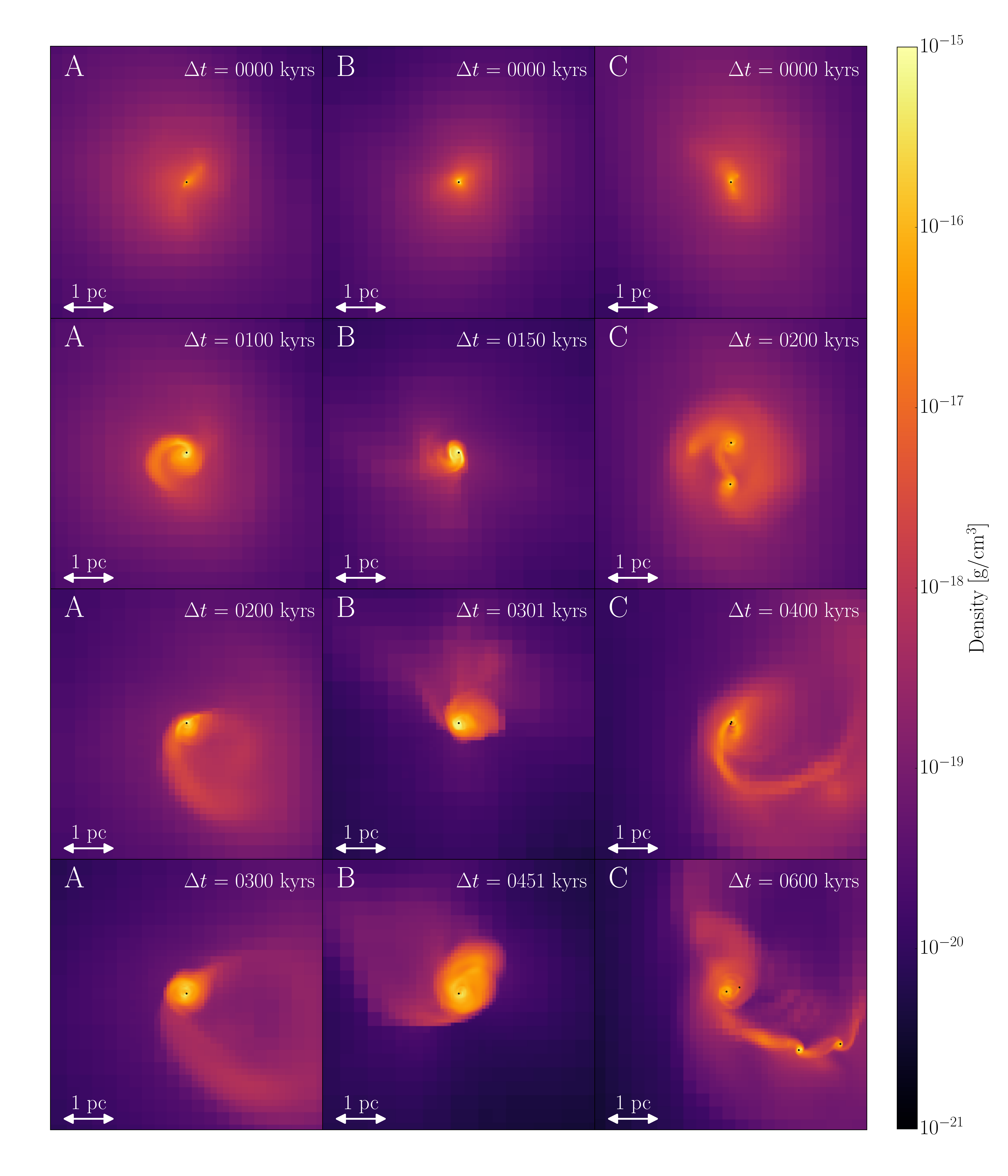}
\caption{Density projection for all re-simulated halos irradiated by a UV background of $J_{21} = 10000$ at a scale of 5 pc. Halos A, B, and C are shown from left to right, respectively. Several epochs are shown, $\Delta t$ represents the age of the simulation relative to the age of the first sink particle formed in the respective run.}
\label{fig:j21_10000}
\end{figure*}

In Figure~\ref{fig:j21_10000}, we portray density projections at various cosmic times at a 5 pc scale. In all of these simulations, a gas disk is formed, so the projection has been preferred to be in the axis perpendicular to the disk plane. The times shown in the plots ($\Delta t$) are the times relative to the first sink particle formed in the respective simulation (see Table~\ref{tab:halo_sink}). We based these $\Delta t$ values on the growth time and spanning the dynamical evolution for the available outputs. Each black dot represents a sink particle. We can now observe the structure evolution with more detail. Halos A and B form a structure that initially starts with a spherical shape but soon becomes a disk with spiral arms from which matter is accreted; in both cases just one sink particle forms. On the other hand, the evolution of the central structure is different in Halo C than in the others: it begins as a spherical dense core where the first sink particle forms, similar to A and B. However, a second core arises after 100 kyr, from this structure a new sink particle is created, which forms a binary system with the first sink particle. Then, the system tightens as time goes on. At $\Delta t = 400$ kyr we observe the particles in the binary system to be very close and the formation of a dense spiral arm. In this arm new sink particles are created. Finally, the binary system merges at a $\Delta t = 540$ kyr, while the other particles fall to the center in the fragmented arm as we can see at $\Delta t = 600$ kyr.

Different mechanisms may trigger the formation of sink particles. However, a stability Toomre analysis allows us to know if they are formed due to local gravitational instabilities. The Toomre parameter \citep{toomre} is defined in Equation~\ref{ec:toomre_1}, where $\kappa$ is the epicyclic frequency, $c_{s}$ is the sound speed, and $\Sigma$ is the surface density.

\begin{equation}
    Q = \frac{\kappa c_s}{\pi G \Sigma}
    \label{ec:toomre_1}
\end{equation}

Toomre parameter values above unity represent stable regions, while values lower than one represent unstable regions, hence, regions that will collapse. Epicyclic frequency  $\kappa$ is shown in Equation~\ref{ec:kappa}, where $\Omega(R)$ is the angular velocity, and $R$ is the distance, both calculated from a given center.

\begin{equation}
    \kappa^2(R) = 4 \Omega^2 + R \frac{d \Omega^2}{d R}
    \label{ec:kappa}
\end{equation}  

Epicyclic frequency is known theoretically to be constrained between unity and two \citep{binney}. This fact allows us to use 2$\Omega$ instead of $\kappa$ resulting in the form presented in Equation~\ref{ec:toomre_2}. We use 2$\Omega$ instead of $\Omega$ to compensate the uncertainties in $\Sigma$ that arise when we use this parameter in a not idealized context.

\begin{equation}
    Q = \frac{2\Omega c_s}{\pi G \Sigma}
    \label{ec:toomre_2}
\end{equation}

\begin{figure}
    \centering
    \includegraphics[width=\columnwidth]{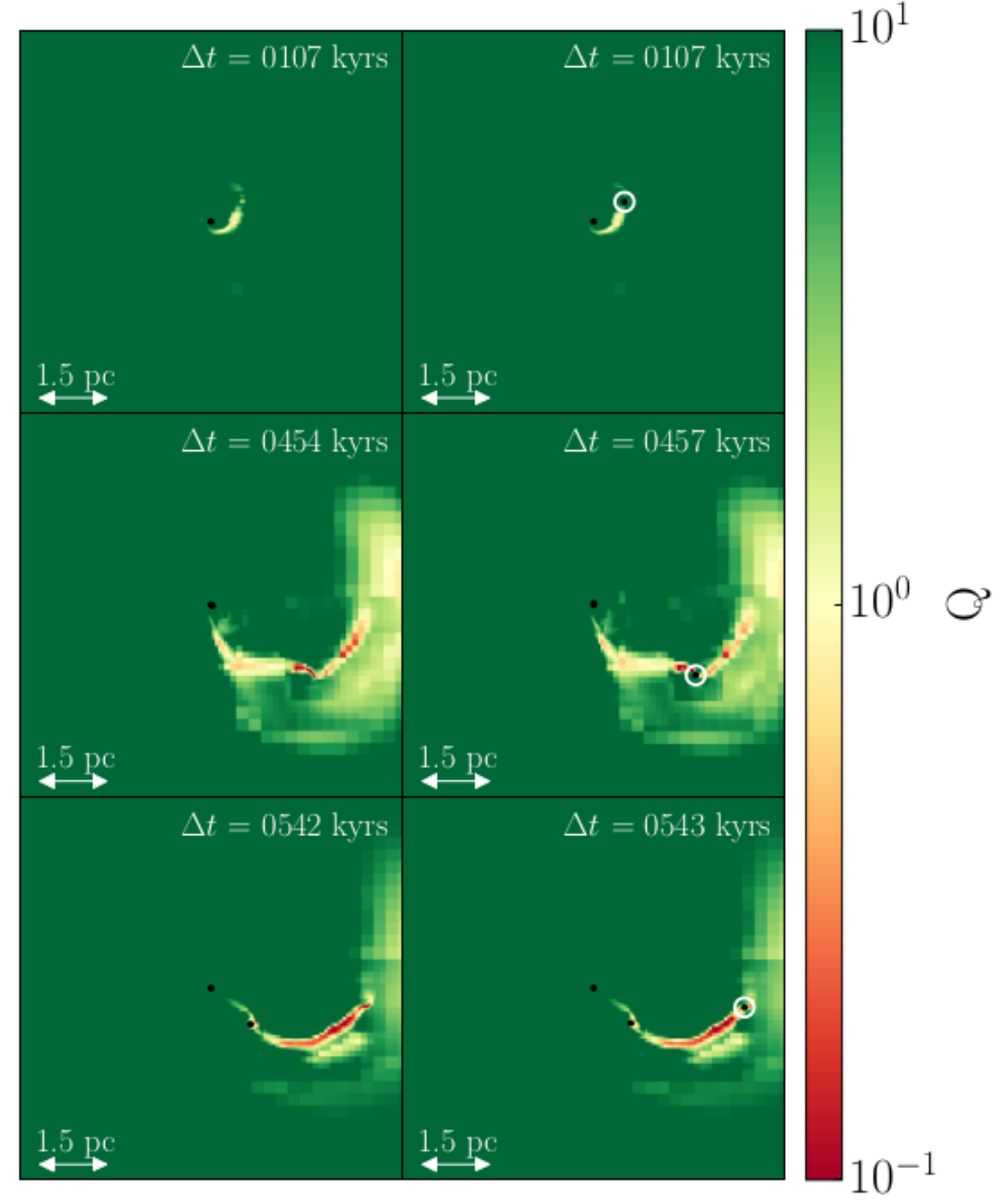}
    \caption{Toomre parameter map for Halo C re-simulations irradiated  by  a UV  background  of $J_{21} = 10000$  at  a  scale  of  7.5 pc, $\Delta t$ are picked one snapshot previous and one later to the formation of three different sink particles. Black dots represent sink particles. $Q$ values higher than one represent stable regions (green), while unstable regions are represented by Toomre parameter values lower than one (red). White circles enclose the region where a sink particle was created.}
    \label{fig:toomre_10000}
\end{figure}

In Figure~\ref{fig:toomre_10000}, we show a Toomre parameter for Halo C at different epochs. We choose this halo due to the larger number of sink particles formed compared to A and B. Unlike previous figures, we show snapshots both before and after sink particle formation. As we can see, regions from which sink particles are made are regions where the Toomre parameter remains much lower than one, implying an imminent collapse. Though this structure is not a perfect thin disk, we can still use it as a check, since most of the structure evolves in plane and the values we found are quite extreme. This simple analysis confirms that sink particles are formed in locally unstable regions.

\subsubsection{Radial Profiles}
\label{sec:rad_prof_10000}

\begin{figure*}
\includegraphics[width=2.1\columnwidth]{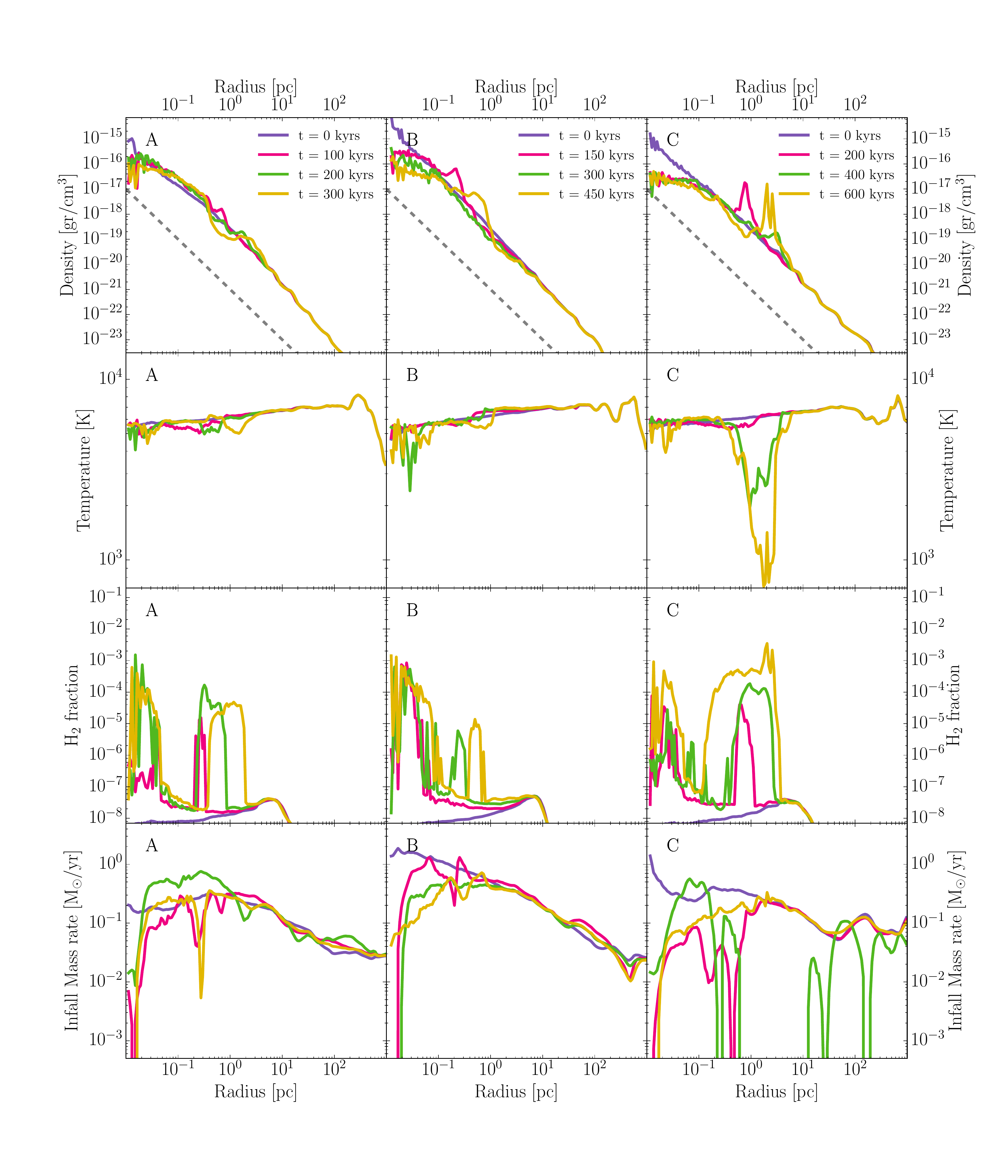}
\caption{Radial profiles for the $J_{21} = 10000$ runs. They include density, temperature, H$_2$ fraction, and infall mass rate. Halos A, B, and C are shown from left to right, respectively. The profiles are calculated at the same times ($\Delta t$) shown in Fig.~\ref{fig:j21_10000}. In density profiles, we plot a dashed gray line that represents an isothermal profile $\rho \propto r^{-2}$. }
\label{fig:profile_10000_v}
\end{figure*}

In addition to the density projections, in Figure~\ref{fig:profile_10000_v} we show radial profiles for these simulations. We include density, temperature, H$_2$ fraction, and mass infall rate. The profiles are also plotted for the same $\Delta t$ we use in Figure~\ref{fig:j21_10000}. All the plots are centered on the first sink particle formed and they span from resolution scales (0.01 pc) up to 1 kpc, i.e. 5 orders of magnitude in spatial scales. In these re-simulations, the virial radii range from 0.9 to 1.5 kpc at the end of the runs, so we are sampling the whole halos.

All density profiles exhibit a nearly isothermal profile $\rho \propto r^{-2}$ (gray dashed line in the top row) with some deviations. For Halos A and B there are some peaks in the profiles that represent spiral arms and the extent of dense regions, as can be seen in the density projections in Figure~\ref{fig:j21_10000}. For Halo C, several peaks appear at different times. The peak at $\Delta t = 200$ kyr represents the second core that forms a binary system. The density profile for $\Delta t = 400$ kyr shows the arise of a small peak, which represents the formation of the spiral arm that can be seen in the density projection. Finally, at $\Delta t = 600$ kyr two peaks appear, that represent the two overdense clumps with sink particles in them.

The temperature remains roughly constant for Halos A and B, but for Halo C we see how it drops down to $\sim$1000 K at some radii. For Halos A and B, the temperature decreases at high densities slowly due to the Lyman-$\alpha$ cooling. This slope has been observed in other works, e.g. \citet{latif2013}. For Halo C at $\Delta t = 0$ and 200 kyr, the temperature profiles are nearly constant, but at $\Delta t = 400$ and 600 kyr, the profile changes drastically in the radii that enclose the spiral arm observed.

\begin{figure}
\includegraphics[width=\columnwidth]{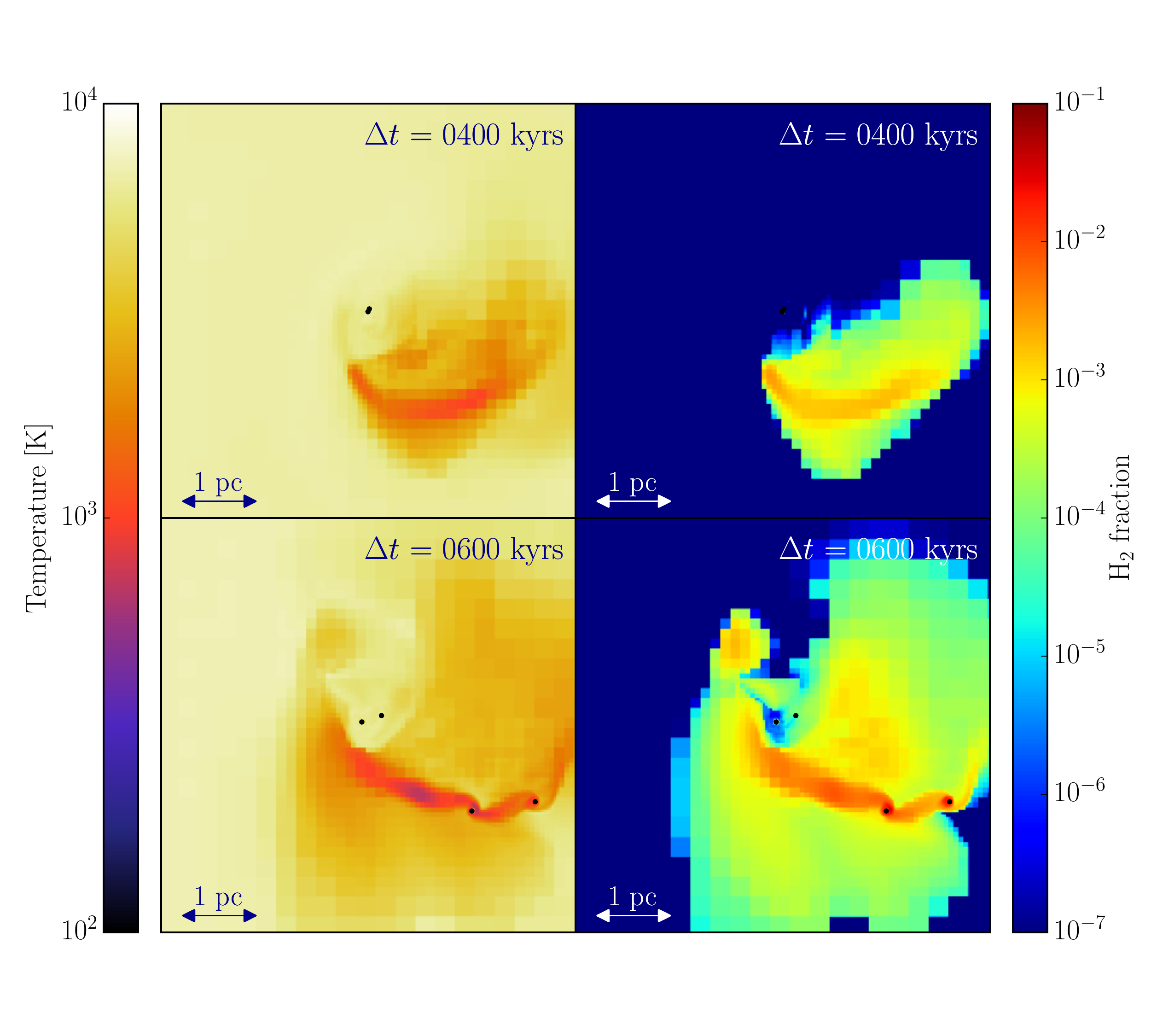}
\caption{Temperature and H$_2$ fraction projections for Halo C re-simulations irradiated by a UV background of $J_{21} = 10000$ at a scale of 5 pc. Temperatures are shown on the left, while H$_2$ fractions are represented on the right. Several epochs are shown, $\Delta t$ represents the age of the simulation relative to the age of the first sink particle formed in the respective run.}
\label{fig:tem_h2}
\end{figure}

The H$_2$ fraction profile for A and B shows some localized H$_2$ enhancement with time, but in both halos, this effect does not lead to a major change in the temperature profiles. This is because the fraction does not exceed a value of 10$^{-3}$. At roughly such a value, cooling becomes efficient lowering the central temperatures. For Halo C there is a huge increase in the H$_{2}$ fraction in the arm formed at the later stages of this simulation, however, the radial profile does not portray this enhancement properly. In Figure~\ref{fig:tem_h2}, we show projections at two different times for the temperature and the H$_2$ fraction. In the left panels, we plot temperature, while in the right ones we plot the H$_2$ fraction. We observe that molecular hydrogen is gathered in the spiral arm seen in the density projection. The increase in H$_2$ also cools down the system in the regions where this molecule is abundant. 

In addition to density, temperature, and H$_2$ fraction, we also added mass infall rates to the radial profiles. The mass infall rates are calculated using Equation~\ref{eq:rate}, centered on the position of the first formed sink particle. Equation~\ref{eq:rate} counts all cells at a given radius and calculates the amount of matter crossing a determined area at a given velocity. In Equation~\ref{eq:rate}, $\rho_{\text{cell,i}}$ represents cell density, $v_{\text{infall,i}}$ corresponds to cell speed along the radial axis towards the center chosen, and $\Delta \Omega_{\text{cell,i}}$ is the solid angle subtended from the center to the portion of the cell that is contributing to the calculations. Notice that $v_{\text{infall,i}}$ has been chosen to be positive if the cell is falling to the center and negative if it is moving away from it.

\begin{equation}
\dot{M}(r) \approx \sum_{i=0}^{ncells} r^2 \rho_{\text{cell,i}} v_{\text{infall,i}} \Delta \Omega_{\text{cell,i}}
\label{eq:rate}
\end{equation}

In all of the runs we can observe regions in where the gas is falling to the center and regions where the gas is moving away. At most radii, infall mass rate calculations turn out to be positive, implying that most of the gas is being accreted compared to gas being pushed away. But there are some regions where most gas is being pushed away, which creates gaps in the mass infall rate profiles.

At some radii we can observe similar behaviors regardless of the halo, this is because $\dot{M}(r)$ scales with $r^2$, $\rho$, and $v_{\text{infall}}$. At high radii, the area considered increases highly, but density and infall velocities are small compared to their values close to the center. On the other hand, the areas taken into account close to the center are small compared to the ones at high radii, but infall velocities are larger. At $\Delta t = 0$ kyr for all runs, the density has the highest values in the center, which makes infall mass rates also have their highest values at those radii and that time. When considering later times, the density values close to the center decrease, which affects the value of the mass infall rate. 

It is interesting to address the fact that mass infall rates are very high in Halos A and B, having infall rates $>$ 0.1 $M_{\odot}$ at different times, while Halo C starts with accretion rates comparable to A and B, but at later times the formation of new sink particles and the interaction between them alters the infall mass rate, reducing the value and generating outflows. This behavior is observed in the central 5 pc as can be seen in Figure~\ref{fig:j21_10000}.

\begin{figure*}
    \centering
    \includegraphics[width=2.1\columnwidth]{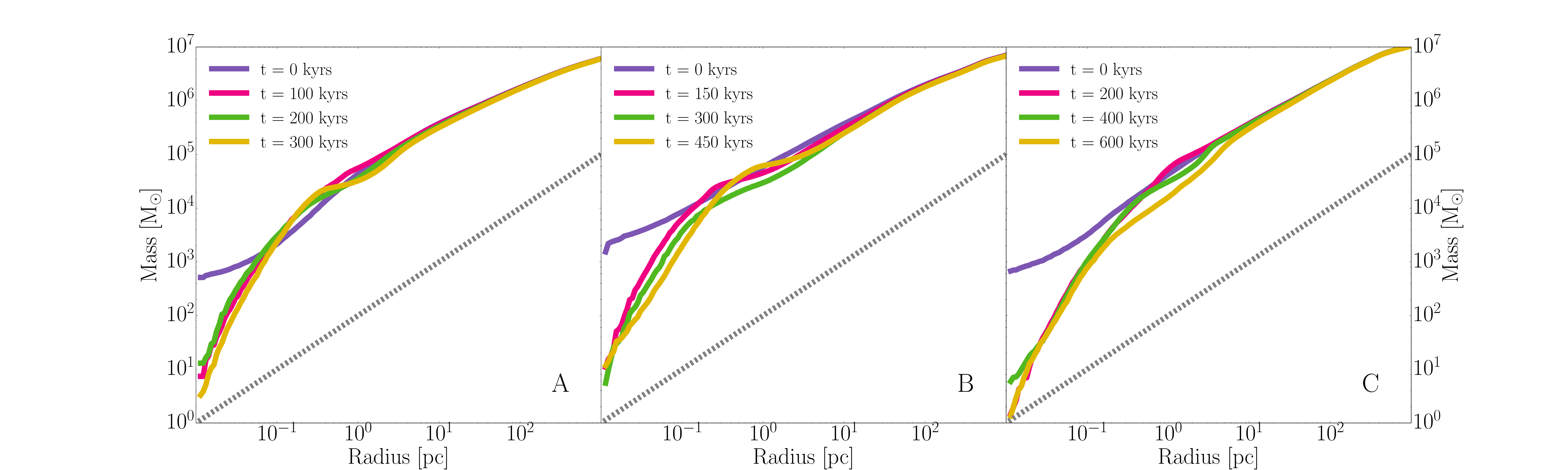}
    \caption{Enclosed mass as a function of radius. The profiles are centered in the position of the first sink particle created in each simulation. Halos A, B, and C are shown from left to right, respectively. The gray dashed line represents the function $M \propto r$.}
    \label{fig:mass_10000}
\end{figure*}

Additionally, we add the enclosed mass profile in Figure~\ref{fig:mass_10000} for all these halos at the same times used in Figure~\ref{fig:profile_10000_v}. In all the cases we observe that at $\Delta t = 0$, mass follows the function $M \propto r$, corresponding to the isothermal profile. We also see that in all the cases the mass drops due to the fact that gas is being accreted in the surroundings. Small bumps appear in the profiles that correspond to overdense regions that can be identified in the density profile.

\subsubsection{Sink particles}
\label{sec:sink_10000}

\begin{figure*}
\includegraphics[width=2.1\columnwidth]{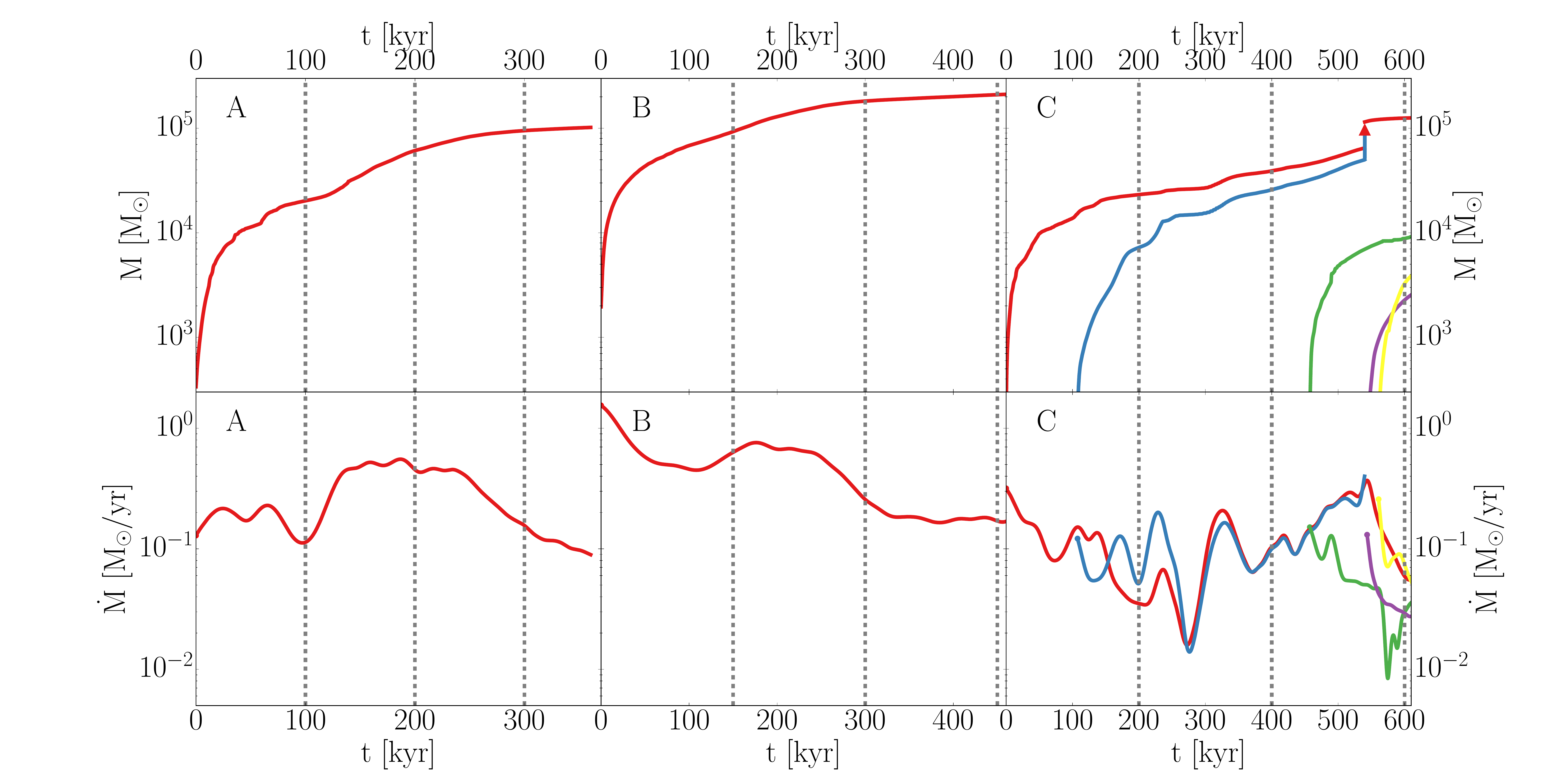}
\caption{Mass and accretion rate time evolution for the $J_{21}$ = 10000 runs. Halos A, B, and C are shown from left to right. Each color represents a single sink particle, the time in the $x$-axis is the time relative to the first sink particle formed in the respective run. Arrows represent merger in the mass profiles, they are bi-colored and start from the end of the mass profile of the least massive sink in the pair that is merging and head to the new profile for the resulting sink particle. The color of the tail of the arrow is the same as the one used for the least massive particle in the pair, while the head has the same color as the most massive sink particle. In mass and accretion rate profiles, the resulting new sink particle maintains the color used for the most massive sink particle in the pair. Accretion rate evolution has been smoothed to follow the long-term evolution.}
\label{fig:sink_10000}
\end{figure*}

As mentioned previously, all points in Figure~\ref{fig:j21_10000} represent sink particles, which interact with other elements in the simulation. Figure~\ref{fig:sink_10000} shows the time evolution for all sink particles in the $J_{21} = 10000$ runs. We show the time evolution of the mass and accretion rate: dashed gray lines represent the times that have been used to picture the projection plots (Figure~\ref{fig:j21_10000}) and to calculate the radial profiles (Figure~\ref{fig:profile_10000_v}). Each color represents a single sink particle and the time in the x-axis is the time relative to the first sink particle formed in the respective run. As mentioned in section~\ref{sec:sink}, sink particles are allowed to merge. These events are represented with arrows in the mass profiles, they are bi-colored and start from the end of the mass profile of the least massive sink in the pair that is merging and head to the new profile for the resulting sink particle. The color of the tail of the arrow is the same as the one used for the least massive particle in the pair, while the head has the same color as the most massive sink particle. In mass and accretion rate profiles, the resulting new sink particle maintains the color used for the most massive sink particle in the pair.

We clarify that the accretion rates for all halos have been smoothed to emphasize long-term variations and variations between sink particles. The smoothing was performed by averaging the accretion rates at a given time with its previous and next value (except by the boundaries). The weights were 0.6 for the central time and 0.2 for each neighbor. The accretion rates shown went through this process for 1000 iterations. This process erased all tiny variations in short time scales, which are associated with gas dynamics around the particles.

From Figure~\ref{fig:sink_10000}, we can see how sink particles evolve with time, observing a huge increase in the masses of the sink particles. In Halo A, the sink particle ends up with 1.0 $\times$ 10$^5$ $M_{\odot}$ near 300 kyr after its formation. The accretion rate remains above 0.1 $M_{\odot}$yr$^{-1}$ during the run, where slight variations within this order of magnitude are observed. For Halo B we find a similar behavior compared to A; there is a huge increase in the mass of the sink particle, ending up with 2.2 $\times$ 10$^5$ $M_{\odot}$. In this case, the accretion rate also remains above 0.1 $M_{\odot}$yr$^{-1}$ but, unlike A, the accretion rate starts at a higher rate $\sim$1 $M_{\odot}$yr$^{-1}$, due to the more massive initial sink mass. Soon it decays slightly, but it maintains a value higher than 0.1 $M_{\odot}$yr$^{-1}$ during the run. This sink particle reaches 10$^{5}$ $M_{\odot}$ in 161 kyr after its formation.

In Halo C, five sink particles are formed. The first forms in a similar environment compared to A and B: a spherical cloud that collapses gravitationally. However, several instabilities arise leading to the formation of a dense core that forms a binary system with the first sink particle. Also, a dense cold arm forms from which new sink particles have created that approach to the center. The first sink particle (red) formed accretes at a rate higher than $>0.1$ $M_{\odot}$yr$^{-1}$ at the beginning. This value oscillates around $0.1$ $M_{\odot}$yr$^{-1}$, and never drops below 0.01 M$_{\odot}$yr$^{-1}$. All other sink particles formed also start with an accretion rate $> 0.1$ $M_{\odot}$yr$^{-1}$ which decrease quickly. All accretion rates are kept over $0.01$ $M_{\odot}$yr$^{-1}$ during the run.

\subsection{$J_{21}$ = 10}

In addition to the run with $J_{21} = 10000$ we also perform simulations with $J_{21} = 10$, expecting fragmentation and the formation of sink particles in order to study their growth.

\subsubsection{Dynamics}
\label{sec:j21_10_dynamics}

We perform simulations for each halo using a uniform UV background of $J_{21}$ = 10. We follow their evolution for  
a larger period of time compared to  the $J_{21} = 10000$ runs. We follow them by $\sim$3 Myr after the first sink particle formed in their respective run.

\begin{figure*}
\includegraphics[width=2.1\columnwidth]{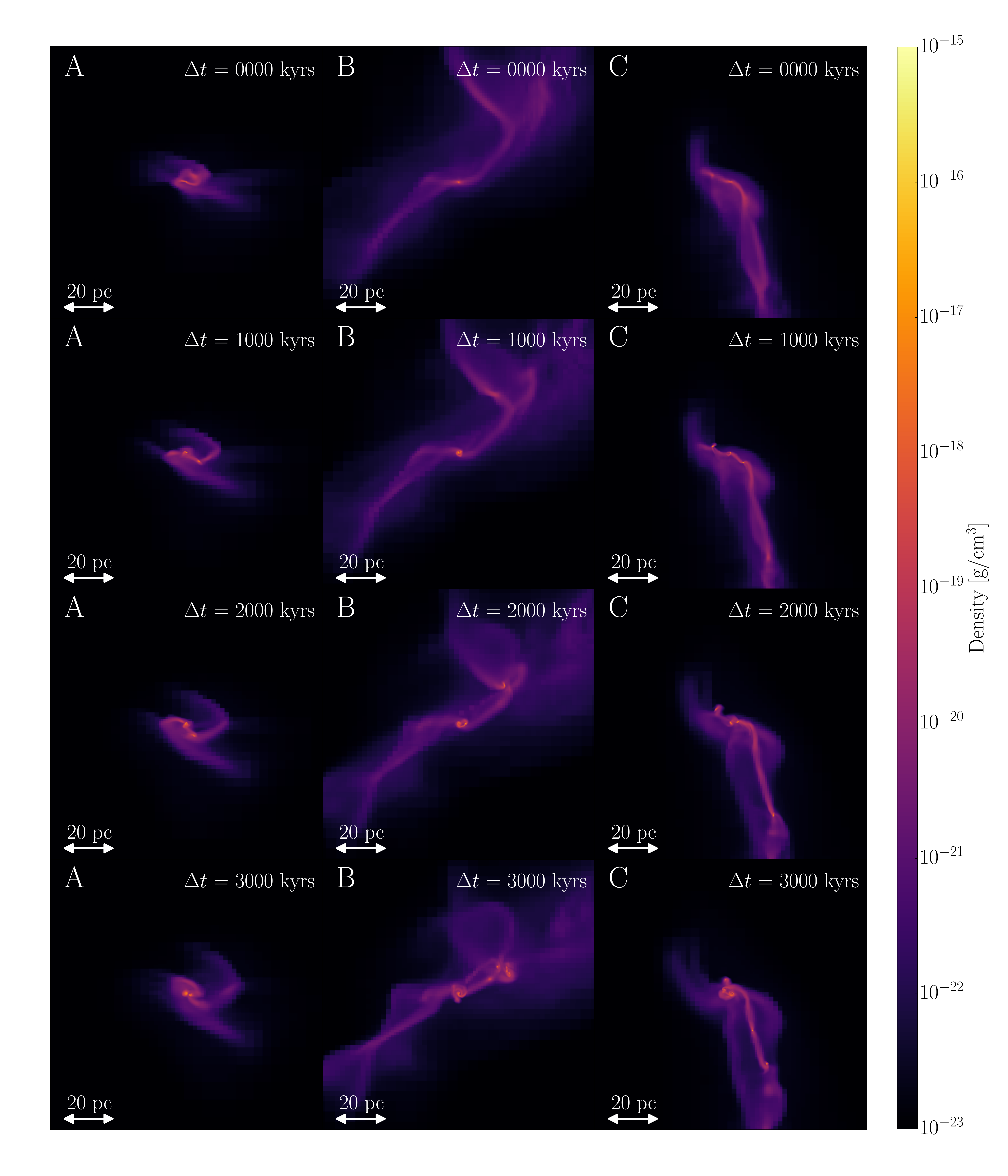}
\caption{Density projection for all re-simulated halos irradiated by a UV background of $J_{21} = 10$ at a scale of 100 pc. Halos A, B, and C are shown from left to right respectively. Several epochs are shown where $\Delta t$ represents the age of the simulation relative to the age of the first sink particle formed in the respective run.}
\label{fig:j21_10_v_100}
\end{figure*}

\begin{figure*}
\includegraphics[width=2.1\columnwidth]{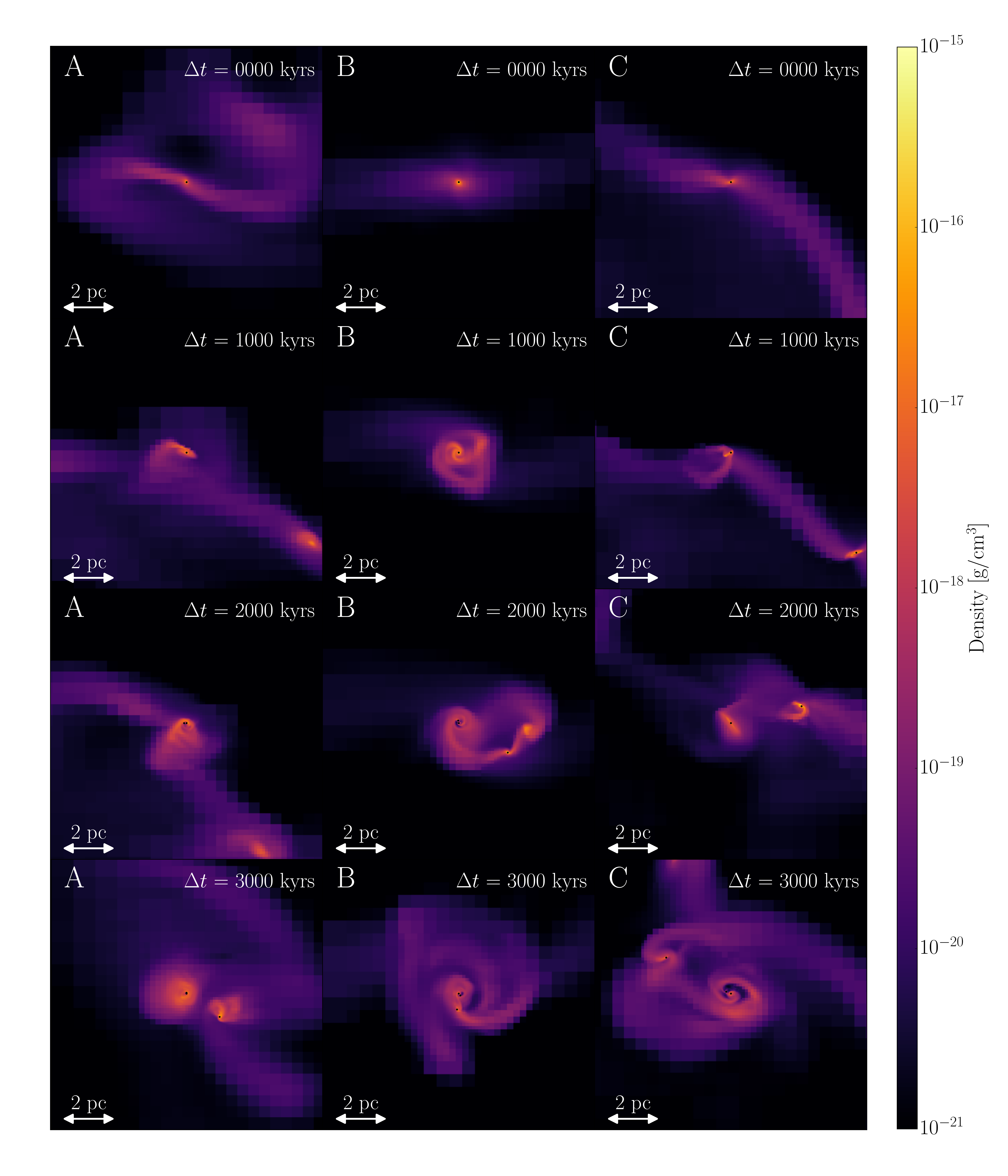}
\caption{Density projection for all re-simulated halos irradiated by a UV background of $J_{21} = 10$ at a scale of 10 pc. Halos A, B, and C are shown from left to right respectively. Several epochs are shown where $\Delta t$ represents the age of the simulation relative to the age of the first sink particle formed in the respective run.}
\label{fig:j21_10_v}
\end{figure*}

In Figures~\ref{fig:j21_10_v_100} and~\ref{fig:j21_10_v} we show the density evolution at various cosmic times in the projection of gas density. This time more than one structure is formed, so the projection is made along the $z$-axis due to the lack of a preferred plane. Figure~\ref{fig:j21_10_v_100} represents the projection at a scale of 100 pc, while Figure~\ref{fig:j21_10_v} pictures it at a scale of 10 pc. The times shown in the plots ($\Delta t$) are snapshots relative to the first sink particle formation time in the respective simulations (see Table~\ref{tab:halo_sink}). In these simulations, $t_{\text{growth}}$ varies much more widely than in the $J_{21}=10000$ case, so we chose $\Delta t$ times based on our computational resources and to better visualize changes.

In Figure~\ref{fig:j21_10_v_100}, we center the projections in the position of the first sink particle formed due to the lack of a reference as a consequence of the larger amount of structures formed. We observe more structures this time compared to the $J_{21} = 10000$ runs at 100 pc. For Halo A, a disk is formed surrounding the sink particle by the moment it forms. Later, a sort of double disk forms, from which two central cores are identified. One of these cores contains the first sink particle formed in its center. In the neighborhood of this core, a second sink particle is formed by $\Delta t = 1.3$ Myr. Also, in the same neighborhood, a third sink particle arises by $\Delta t = 1.7$ Myr, which quickly merges with the first sink particle created. The remaining particles form a binary system that tightens with time. In the other clump, a fourth sink particle forms at $\Delta t = 2.7$ Myr, residing in its center. For Halo B, we initially observe a central core embedded in a filamentary structure extended along the 100 pc shown in Figure~\ref{fig:j21_10_v_100}. We can see how the filament fragments and forms new sink particles as it approaches to the core. For Halo C, we also see a filament with a central structure. By $\Delta t = 1$ Myr it has developed four cores having sink particles. We also see how the three central ones approach each other with time.

In Figure~\ref{fig:j21_10_v} we portray a 10-pc scale zoom-in from Figure~\ref{fig:j21_10_v_100}, in which every black dot represents a sink particle. This time the structures are so extended that we lose information from the whole halo by zooming-in. {red}{Halo A forms at the beginning a single structure within these 10 pc. By $\Delta = 1$ Myr, we observe a central sink particle and another dense core. Later, and as mentioned previously, two sink particles are formed in its surroundings, but we are not able to see how they interact due to the time sampling used to make the figures. By $\Delta t = 2$ Myr, a merger has already happened and a binary system between the remaining two particles is observed. By $\Delta t = 3$ Myr, we see the central structure with the resulting sink particle in its center, along with the approach of another dense region visible in Figure~\ref{fig:j21_10_v_100} since $\Delta t = 1$ Myr, but it did not form a sink particle until $\Delta t = 2.7$ Myr. When the simulation finishes, the central binary system remains at a distance of 0.03 pc, while this system remains at a distance of $\sim$
5.9 pc to the sink particle in the other core.}

\begin{figure}
\includegraphics[width=\columnwidth]{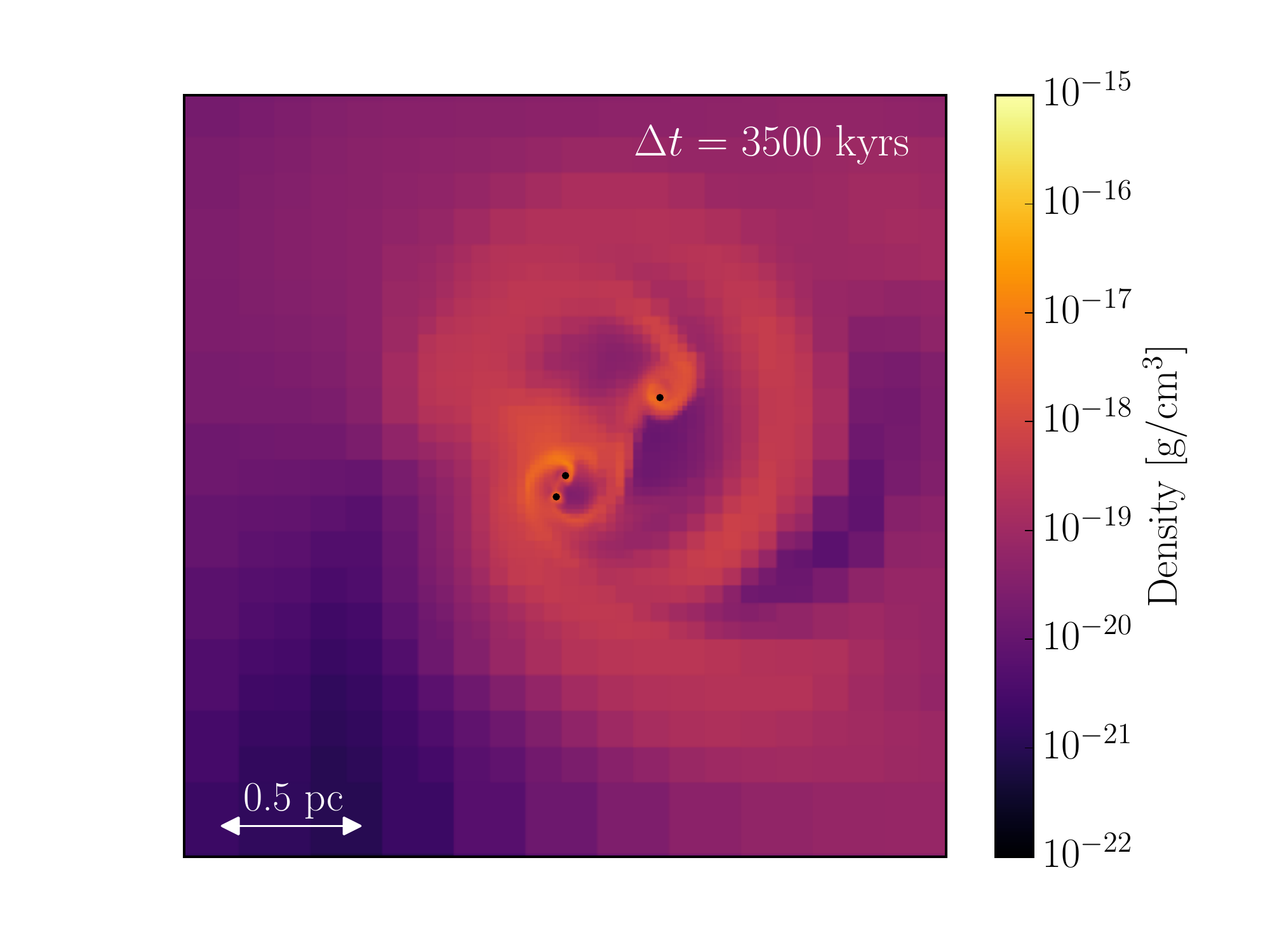}
\caption{Density projection for the central 2.5 pc in Halo B. The projection has been made along the $z$-axis. Black dots represent sink particles. This projection has been made at $\Delta t = $ 3.5 Myr after the formation of the first sink particle in this run.}
\label{fig:j21_10_b}
\end{figure}

In Halo B we can see the formation of a sink particle inside of a spherically shape structure at the beginning. Later, this sphere becomes a disk-like structure with the sink particle in its center at $\Delta t$ = 1 Myr. After that, a second sink particle is created in the disk surroundings at $\Delta t$ = 1.1 Myr. This particle falls down to the center, but instead of merging with the central one, they begin to orbit each other forming a binary system. The binary remains stable, and a density gap between them arises \citep{luciano_2012,luciano_2014,luciano_2015}. Later, another sink particle is formed in the 5 pc region, created from a dense region that can be seen at $\Delta t = 2$ Myr, it approaches to the center. This sink particle falls down to the center, merging with the most massive member of the binary. The resulting sink particle forms a new binary system with the left particle. We also distinguish another dense spherical region next to this new sink particle, which forms a new sink particle at $\Delta t = 2.5$ Myr. By $\Delta t = 3$ Myr, we can see the interaction of this new binary with the aforementioned sink particle. This particle forms a triple system with the other particles. In Figure~\ref{fig:j21_10_b}, we show a later stage for this halo ($\Delta t = 3.5$ Myr), where we can see the approach of this sink particle to the center. In this run 13 sink particles are created, but they are so widely distributed that not all of them can be seen in Figure~\ref{fig:j21_10_v}. It is interesting to note that a sink particle forms ~30 pc away from this structure at $\Delta t = 1.6$. This sink particle evolves the other fragmented dense region that can be seen in Figure~\ref{fig:j21_10_v_100}, and it reaches the largest amount of mass after a merger.

In Halo C a single filament is observed from which sink particles are formed. At $\Delta t = 0$ Myr, we observe the first sink particle embedded in a dense isolated region. At $\Delta t = 1$ Myr, we recognize this single particle to be in a disk. By this time, two other sink particles were already formed, but only one is included in Figure~\ref{fig:j21_10_v}. One of these sink particles is 6.6 pc away from the first sink particle, while the other one is at 7.6 pc. Both particles are formed in the same filament observed in Figure~\ref{fig:j21_10_v_100}. At $\Delta t = 2$ Myr, the first sink particle in the center is now seen edge-on, and we observe the approach of one sink particle. By this time, a fourth sink particle has been formed, but at a further distance of $\sim$37 pc from the first sink particle. At $\Delta t = 3$ Myr we observe that the first and the second sink particles have formed a binary system, while a third one approaches to the center, no merger happens during this run.

\begin{figure}
    \centering
    \includegraphics[width=\columnwidth]{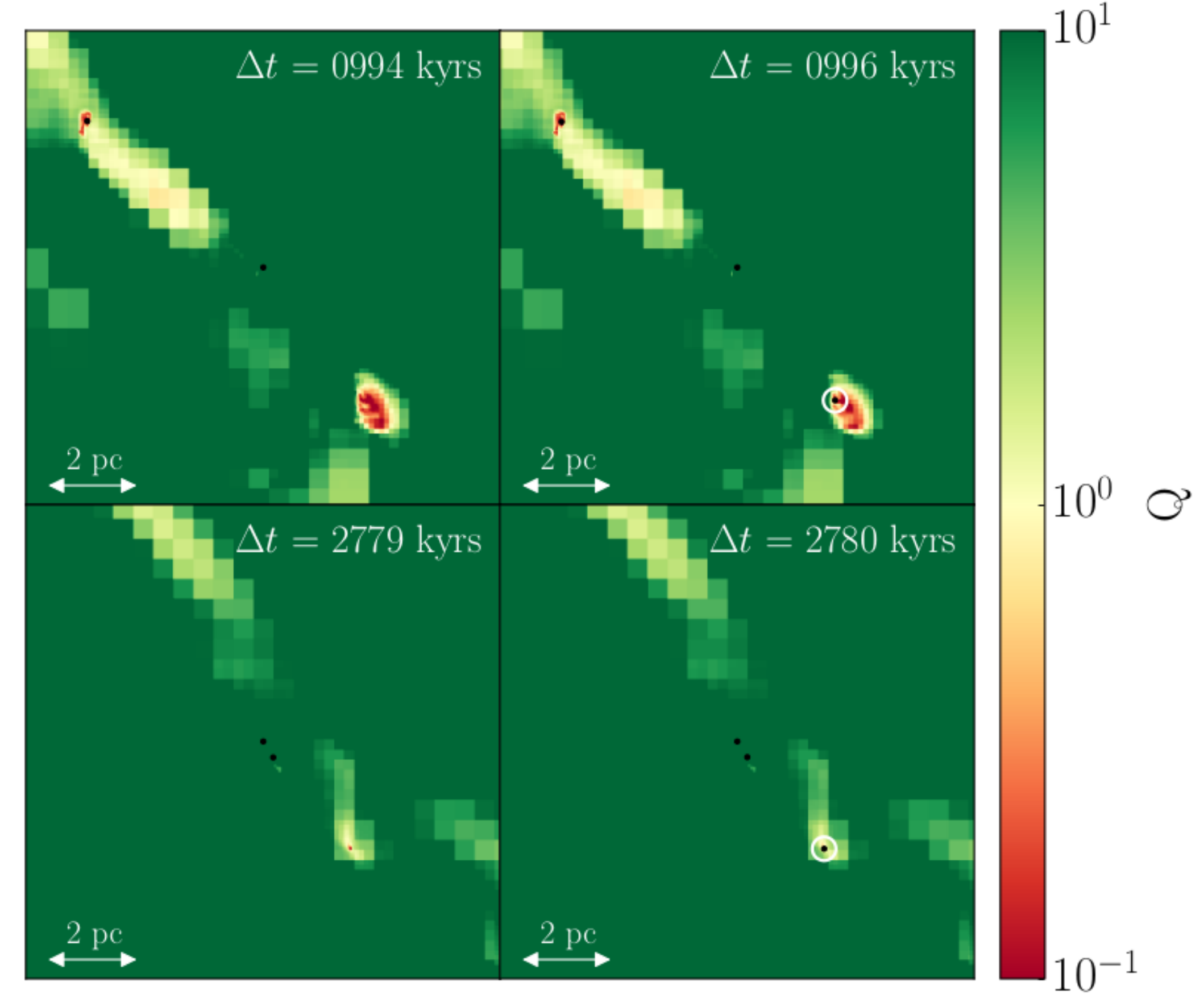}
    \caption{Toomre parameter for Halo C re-simulations irradiated by a UV background of $J_{21}$ = 10 at a scale of 7.5 pc; $\Delta t$ values are picked one snapshot previous and one later to the formation of two different sink particles. Black dots represent sink particles. Values higher than one for $Q$ represent stable regions (green), while unstable regions are represent by Toomre parameter values lower than one (red). White circles enclose the region where a sink particle has been created.}
    \label{fig:toomre_10}
\end{figure}

Similar as we show in Section~\ref{sec:j21_10000_dynamics}, we study the nature of the collapse that leads to the formation of sink particles. In these runs, all halos fragment and form a variety of sink particles. In Figure~\ref{fig:toomre_10}, we show a Toomre parameter map for Halo C at different epochs to illustrate the common condition for the formation of sink particles in our runs. Unlike previous figures, we show snapshots both before and after sink particle formation. As we can see, sink particles form in regions where the Toomre parameter is much lower than one, implying an imminent collapse. This  simple analysis confirms that sink particles are formed in locally unstable regions.


\subsubsection{Radial profiles}

\begin{figure*}
\includegraphics[width=2.2\columnwidth]{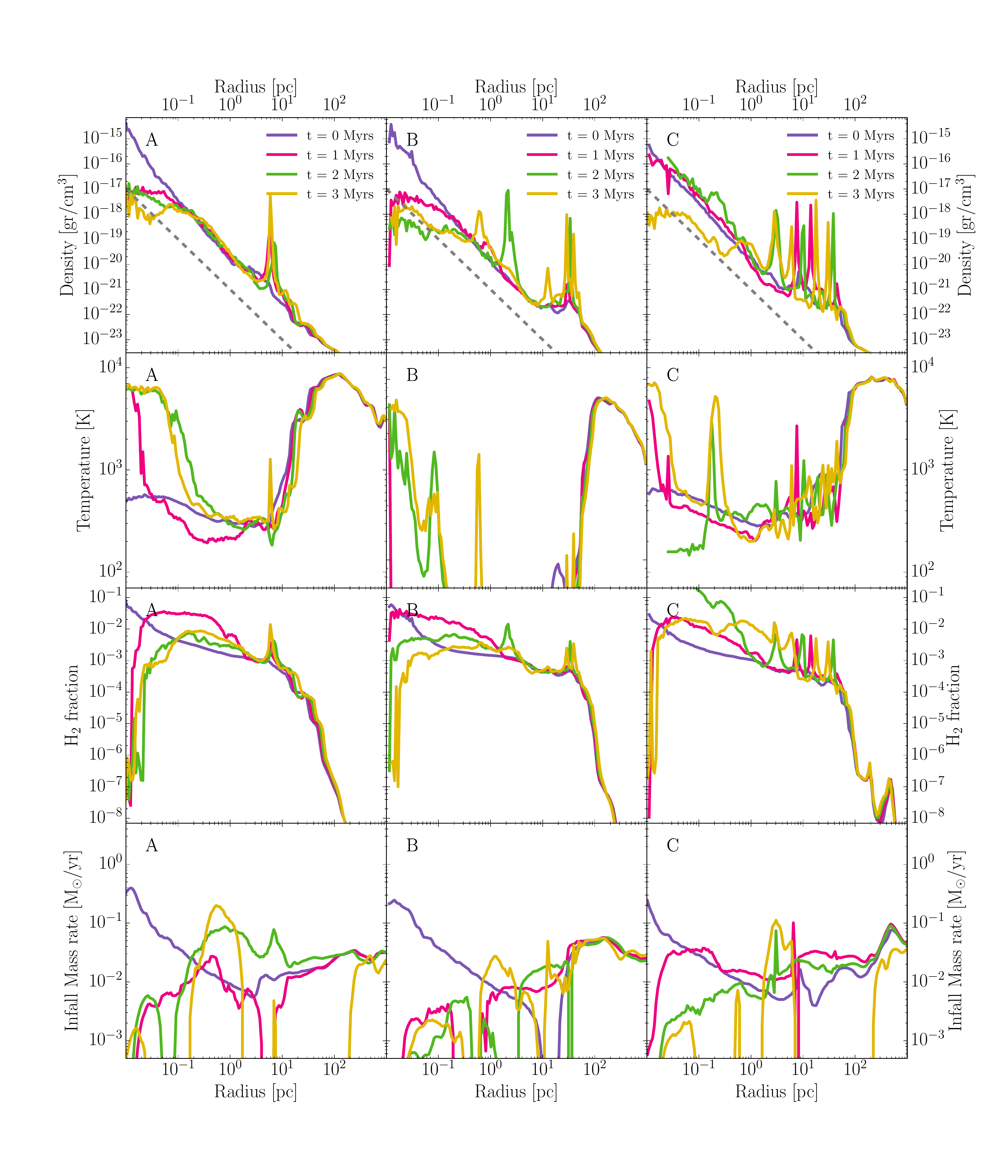}
\caption{Radial profiles for the $J_{21}$ = 10 runs. They include density, temperature, H$_2$ fraction, and infall mass rate. Halos A, B, and C are shown from left to right, respectively. The profiles are calculated at the same times ($\Delta t$) shown in Fig.~\ref{fig:j21_10_v_100}. In density profiles, we plot a dashed gray line that represents an isothermal profile $\rho \propto r^{-2}$. }
\label{fig:profile_10_v}
\end{figure*}

In addition to the density projections, we show radial profiles for these simulations in Figure~\ref{fig:profile_10_v}. We include density, temperature, H$_2$, and mass infall rate. The profiles are also plotted for the same $\Delta t$ we have used in Figures~\ref{fig:j21_10_v_100} and~\ref{fig:j21_10_v}. All of the plots are centered in the first sink particle formed and span from resolution scales $\sim 0.01$ pc up to 1 kpc in order to sample the whole halo. The virial radii range from 0.8 up to 1.1 kpc at the end of the runs. They are smaller compared to the $J_{21} = 10000$ runs, since gravitational collapse occurs at a later cosmic age.

All density profiles exhibit a nearly isothermal profile $\rho \propto r^{-2}$ (gray dashed line) with some deviations. For all halos, there are some peaks in the profiles that represent overdense regions away from the chosen center, some of which host sink particles. In all cases, we observe a decrease in the central density with time due to gas accretion in the neighborhood of the central sink particle.

In Halo A, we visually identified a binary system (see Section~\ref{sec:j21_10_dynamics}), with one of the components being the center of its radial profile. The other component of the binary is observed as a peak in density, which varies in position at different times.

In Halo B, the peaks deviating from the isothermal profile represent overdense regions as seen in Figure~\ref{fig:j21_10_v_100}. At $\Delta t = 0$ Myr, we see a small density peak at $\sim$ $30$ pc, which increases a little by $\Delta t = 1$ Myr. At $\Delta t = 2$ Myr, we appreciate the same peak with a much higher value: this is caused by the collapse of that region, we also observe a peak at $\sim$2 pc that represents the dense regions next to the binary system in Figure~\ref{fig:j21_10_v}. As a consequence of fragmentation, new dense regions are formed. Three peaks are identified between 10 and 30 pc at $\Delta t = 3$ Myr. All density peaks observed in the density profile can be identified in Figure~\ref{fig:j21_10_v_100}.

In Halo C, we found similarities to Halo B in terms of the formation of collapsing structures away from the center. These structures are seen in Figure~\ref{fig:j21_10_v_100} and are also identified by looking at peaks in the density profiles. At $\Delta t = 0$ Myr we observe two slight overdensities at $\sim$ $10$ pc and at $\sim$ $40$ pc.
These are identified as regions that will collapse to form denser structures and sink particles. At $\Delta t = 1$ Myr, we identify three peaks, two are at $\sim$6 pc and $\sim 40$ pc that resemble the slight overdensities mentioned above, while a new peak appears at $\sim$20 pc. Also, a new overdensity is observed at $\sim$14 pc. By $\Delta t = 2$ Myr, density peaks have been displaced to new positions, where sink particles have formed by this epoch.

The temperature varies more widely than in the $J_{21}$ = 10000 case. For the three halos, we observe that temperature is higher than 5000 K at distances larger than $\sim100$ pc from center. This phenomenon tells us about the distance at which interactions and fragmentation are enclosed. Inside this region gas cools down to a few hundred Kelvin and Halo B reaches even lower values. For the three runs we observe that, as time goes on, the central temperature rises. This is due to the dynamical heating produced by the interactions between the central sink particle and its surroundings.

The H$_2$ fraction profile shows a similar behavior compared to the temperature profiles for all halos. It shows a decrease in its value starting from a given distance. This leads to an increase in temperature. Usually, H$_2$ fraction and temperature behaviors are related: high H$_2$ concentrations cool down the system to low temperatures, and high temperatures imply a lack of enough coolants. We appreciate such a trend even at small radii for Halo A. However, we do not observe such a correlation in the central regions of Halos B and C. As mentioned in section~\ref{sec:j21_10_dynamics}, Halo B turns to form a central disk with sink particles orbiting each other in the center forming a binary system. The dynamical interaction between this binary and the gas results into gas heating, which is compatible with high concentrations of molecular hydrogen and high temperatures. Halo C shows a similar behavior to B in terms of temperature and H$_2$ fraction. Though Halo C does not form a tight structure compared to Halo B, their sink particles do interact with each other heating up gas.

In addition to density, temperature, and H$_2$ fraction, we also add mass infall rates to radial profiles; these profiles are calculated in the same way as it was explained in Section~\ref{sec:rad_prof_10000}.

In Halo A, we observe the highest accretion rate $\Delta t = 0$. At $\Delta t = 1$ Myr and $\Delta t = 2$ Myr, we observe a decrease in the mass infall rate, especially in the center, where gas has been accreted by the central sink particle. By $\Delta t = 3$ Myr, we observed the infall mass rate to have many gaps, which represent outflows.

Halo B shows its highest central density at $\Delta t = 0$ Myr, with a gap around $\sim10$ pc. At later times this value decreases, as in all other cases. All of the interactions between sink particles and the formation of binaries (see Section~\ref{sec:j21_10_dynamics}) lead to outflows that appear as gaps in the profile.

In Halo C, we do not observe any gaps at $\Delta t$ = 0, 1, and 2 Myr, since the central structure does not interact strongly with other structures compared to halos A and B. At $\Delta t = 3$ Myr, we identify a gaps that arise from the interaction between the central binary system and the third sink particle that approaches.

\begin{figure*}
    \centering
    \includegraphics[width=2.1\columnwidth]{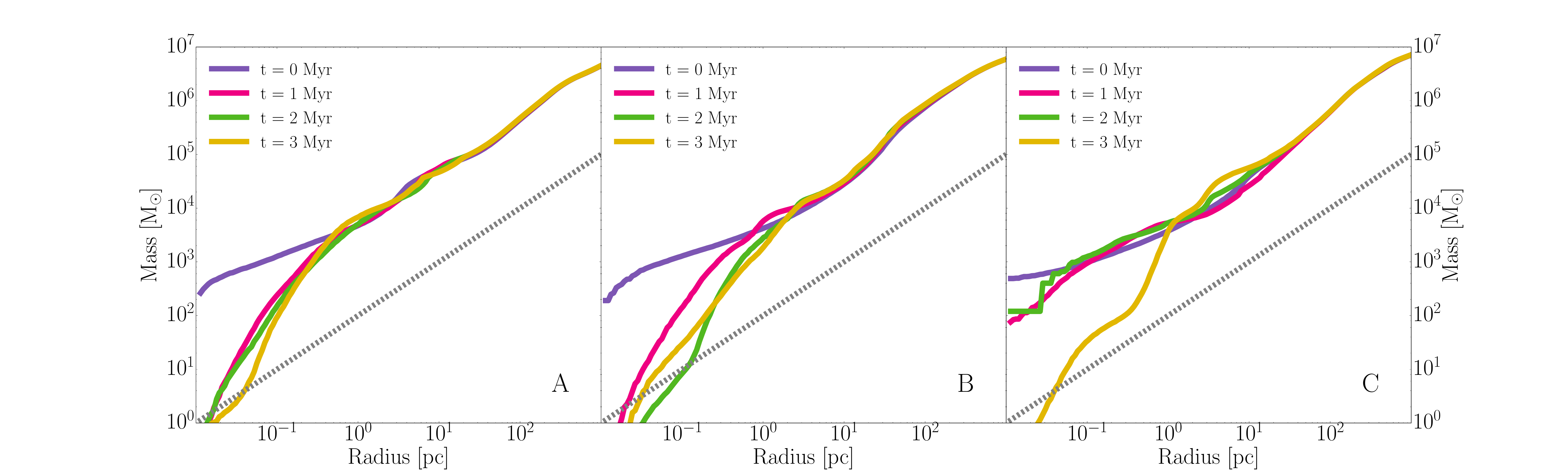}
    \caption{Enclosed mass as a function of radius. The profiles are centered in the position of the first sink particle created in each simulation. Halos A, B, and C are shown from left to right, respectively. The gray dashed line represents the function $M \propto r$.}
    \label{fig:mass_10}
\end{figure*}

Additionally, we add the enclosed mass profile in Figure~\ref{fig:mass_10} for all these halos at the same times used in Figure~\ref{fig:profile_10_v}. For all the cases we observe that at $\Delta t = 0$, mass follows the function $M \propto r$, corresponding to the isothermal profile. We also see that in all the cases the mass drops due to the fact that gas is being accreted in the surroundings. Since the mass of the sink particles in these cases are less than in the $J_{21}$ runs, the accretion rates are as well, and the mass cannot reach the sink particle, leading to a bigger drop in masses at small radii. Bumps appear in the profiles that correspond to overdense regions that can be identified in the density profile.

\subsubsection{Sink Particles}
\label{sec:sink_j21_10}

\begin{figure*}
\includegraphics[width=2.1\columnwidth]{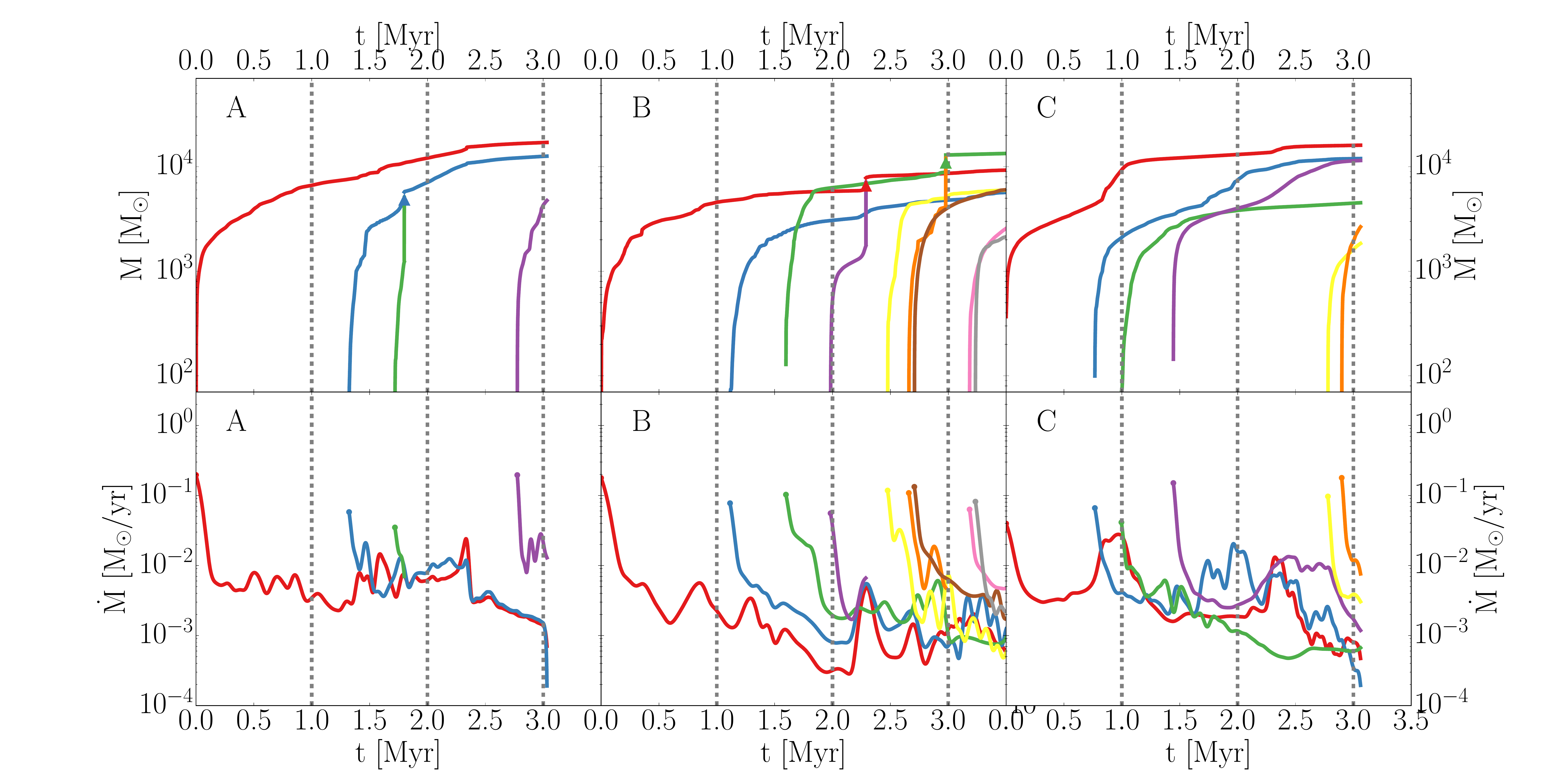}
\caption{Mass and accretion rate time evolution for the $J_{21}$ = 10 runs. Halos A, B, and C are shown from left to right. Each color represents a single sink particle, the time in the $x$-axis is the time relative to the first sink particle formed in the respective run. Arrows represent mergers in the mass profiles; they are bi-colored and start from the end of the mass profile of the least massive sink in the pair that is merging and head to the new profile for the resulting sink particle. The color of the tail of the arrow is the same as the one used for the least massive particle in the pair, while the head has the same color as the most massive sink particle. In mass and accretion rate profiles, the resulting new sink particle maintain the color used for the most massive sink particle in the pair. Accretion rate evolution has been smoothed in order to follow the long-term evolution.}
\label{fig:sink_10_v_1000}
\end{figure*}

In Figure~\ref{fig:j21_10_v} we represent every single sink particle with a black dot. As mentioned previously, not all sink particles lie in the 5 pc shown in the figure. In Figure~\ref{fig:sink_10_v_1000} we show the time evolution for all sink particles created. We show the time evolution of mass and accretion rate: dashed gray lines represent the times that have been used to picture the projection plots (Figures~\ref{fig:j21_10_v_100} and~\ref{fig:j21_10_v}) and to calculate the radial profiles (Figure~\ref{fig:profile_10_v}). Each color represents a single sink particle, the time in the $x$-axis is the time relative to the first sink particle formed in the respective run. Arrows represent merging particles in mass profiles, and accretion profiles have been smoothed for easy reading. Details of both procedures are in Section~\ref{sec:sink_10000}. 

In Figure~\ref{fig:sink_10_v_1000} we can see how sink particles change with time. Their masses increase naturally, reaching values over 10$^3$ $M_{\odot}$. The most massive sink particles reach masses over 10$^4$ $M_{\odot}$. We emphasize that these sink particles take more time to reach a significant amount of mass compared to the $J_{21} = 10000$ case, where they take hundreds of kyr to do it. All sink particles start with accretion rates$~\sim 0.1$ M$_{\odot}$ yr$^{-1}$, whose value decays at least one order of magnitude in all cases. As fragmentation raises in these runs, regions where sink particles are created do not contain enough gas to reach initial accretion rate values compared to the $J_{21} = 10000$ runs. We also observe a rapid decrease in accretion rates with time, which is mainly due to the fact that the gas has been accreted. For Halo A, we observe some oscillatory behavior in the accretion rate, generated by the binary interaction we see in the density map. For Halo B, we identify a succession of binaries and sink particle mergers. The participants are represented in Figure~\ref{fig:sink_10_v_1000} in blue, red, purple, and yellow. We also observe oscillatory accretion rates. In Appendix~\ref{sec:accretion} we show some accretion rates with no application of smoothing technique in the binaries regime for Halos A and B.


In Halo C, the situation is a little bit different than in the previous cases, mainly due to the fact that sink particles are spread over 100 pc, and we see a few interactions.

\section{Discussion}

In this work, the formation of supermassive black hole seeds has been studied including the role of fragmentation processes. The DCBH scenario seems very efficient forming massive seeds ($> 10^{5}$ $M_{\odot}$), which become massive black holes when the general relativistic (GR) instability \citep{chandra} is triggered. Such black holes may end up in the center of the high redshift quasars observed. However, it demands several strong restrictions in order to fulfill its assumptions. Between these restrictions, strong gas fragmentation is not allowed, since it contradicts the aim of forming a central single massive structure.

We performed 3D cosmological simulations in order to explore how gas behaves under different conditions, including DM halo spin, DM halo merger history, and background UV intensity. Two of three re-simulated halos under the presence of a high UV intensity ($J_{21} = 10000$), namely A and B, replicate very well the DCBH scenario. They correspond to the halos with the lowest spin parameters and also to halos formed out of a small number of mergers. These two features, i.e. halo spin parameter and merger history, seem to be relevant for the formation of SMBH seeds. These halos resemble the results reported in other works \citep{latif2013,latif2013a}, in which a central core is formed without any fragmentation. In these simulations, just one sink particle is formed in each simulation during the whole run. The radial profiles for them support the formation of a central massive structure: density profiles follow nearly an isothermal profile, temperatures are kept roughly constant, H$_2$ fractions do not cool the system down, and mass infall rates are in the range required to ensure a relevant growth.

On the other hand, in the re-simulation for Halo C, a different behavior is observed. It is recalled that Halo C is the DM halo with the highest spin parameter from the ones chosen and formed out the interaction of a considerable amount of mergers. This halo starts in a similar fashion to A and B; however, as the simulation goes on, fragmentation processes appear in the central 5 pc, which is seen as a disk-like structure with a spiral arm from which sink particles are formed. Sink particles formed in this case are still massive enough with values close or higher than 10$^4$ $M_{\odot}$. One sink merger is observed and, according to the behavior of the sink particles, all could end up merging. In such a case, the final mass would be higher than 9.5$\times 10^{4}$ $M_{\odot}$. One of the most interesting phenomena of this run is the high concentration of molecular hydrogen as can be seen in the radial profiles and H$_{2}$ fraction projection for this.

In addition, three re-simulations were performed for Halos A, B, and C under the presence of a weak UV background ($J_{21}$ = 10). The evolution of these structures is much more different than in the other cases since fragmentation is observed, and at larger scales. Halo A is an interesting case as fragmentation is observed in the central 20 pc and it remains at that scale during the whole run. In this case, two sink particles are formed in two core regions, they evolve independently to get bound at the end of the run, but do not merge. By the time the run finishes, their masses are 2$\times10^{4}$ $M_{\odot}$ and 4$\times10^{3}$ $M_{\odot}$. In Halos B and C, the fragmentation process is seen to be extended in 100 pc, the number of sink particles is the largest for all the simulations performed being 8 and 7 for Halos B and C, respectively. Also, in both halos, the mass of the sink particles is higher than 10$^{3}$ $M_{\odot}$ at the end of the runs. An important fact is that sink particles tend to approach to the most massive one with time, a few interactions and mergers are seen. Interestingly, in all the of simulations, the first sink particles reach similar masses, being close to 10$^{4}$ $M_{\odot}$, while the ones formed at later times keep $\sim10^{3}$ $M_{\odot}$ as can be seen in the sink evolutionary profiles.

Several works have tried to find the critical UV intensity value ($J_{21}^{\text{crit}}$) at which a DCBH can take place \citep{latif2015c,glover2015}. Finally, from these results, $J_{21}^{\text{crit}}$ seems to be a parameter that depends critically in the morphology of a particular system rather than anything else. Halos A and B fragment under the presence of a UV intensity of $J_{21}$ = 10, while they form a single massive object under the presence of a UV intensity of $J_{21}$ = 10000. On the other hand, in our halo C, we see fragmentation in both cases and if we include the halo C under the presence of UV intensity of $J_{21}$ = 100 (Appendix~\ref{sec:j21}). It seems that it is not possible to parameterize the phenomenon with the $J_{21}$ parameter only, since the dynamics of the system play a key role in the formation of SMBH seeds.

One of the caveats of the simulations performed is that they do not include radiative feedback in their implementation. It is known that stars radiate photons, supporting gas from being accreted into a black hole seed. \citet{hosokawa2012,hosokawa2013} studied the evolution of rapid accreting supermassive stars (SMS). They found that accreting SMS at rates higher than $> 0.1 M_{\odot}$yr$^{-1}$ evolve as supergiant stars in which relativistic instability is triggered when they reach masses $\sim10^{5}$ $M_{\odot}$; they also found that ionizing radiation is unlikely to operate. These findings are compatible with the analytical results found by \citet{dominik13}. More recently, \citet{haemmerle} studied accreting protostellar evolutionary models in a similar way to the previous works mentioned obtaining similar results. In their work, most massive stars evolve as red and cold supergiant stars, while the less massive ones evolve toward the ZAMS as blue and hot stars. They reduced the critical accretion rate at which ionizing radiation becomes negligible down to 0.005 M$_\odot$/yr. In addition, \citet{chon2018} performed radiative hydrodynamical cosmological simulations to study the effect of radiation in the growing of the SMS, finding no major role of radiative feedback.

The requirements to keep radiation unimportant according to the conditions mentioned in the works above are met in all the simulations performed at high UV intensity. Though halo C fragments, all sink particles formed in this simulation meet the minimum accretion rate requirement and the masses to trigger the GR instability. 

From an observational point of view, \citet{pacucci2016} devised a method to identify supermassive black hole seed candidates based on the spectra of radiative hydrodynamic simulations in the DCBH scenario. They claimed findings of two possible candidates, using infrared and X-ray photometry. The \textit{James Webb Space Telescope}\footnote{\href{https://www.jwst.nasa.gov/}{https://www.jwst.nasa.gov/}} is alleged to be a key instrument to obtain spectra from these objects unveiling clues in the understanding of the mechanism to form SMBHs. The recent detection of gravitational waves (GWs) from the merging of a BH binary by LIGO \citep{ligo} has opened a new area of study. If the sink particles formed are considered relativistic objects, their interaction would emit detectable GWs. GW were speculated by \citet{rees_1978} in this context. \citet{pacucci_grav} showed that the gravitational signal in the context of an asymmetric collapse \citep{shapiro_2003} lies above the foreseen sensitivity of the Ultimate-DECIGO observatory in the frequency range (0.8–300) mHz.

\acknowledgments

We received partial support from Center of Excellence in Astrophysics and Associated Technologies (PFB-06). M.S., J.P., and A.E. acknowledge partial support from FONDECYT through grant 1181663. D.R.G.S. acknowledges partial support from FONDECYT through grant 1161247 and CONICYT PIA ACT172033. The Geryon cluster at the Centro de Astro-Ingenieria UC was extensively used for the calculations performed in this work. BASAL CATA PFB-06, the Anillo ACT-86, FONDEQUIP AIC-57, and QUIMAL 130008 provided funding for several improvements to the Geryon cluster. Powered@NLHPC: This research was partially supported by the supercomputing infrastructure of the NLHPC (ECM-02). We use the toolkit yt\footnote{\href{http://yt-project.org/}{http://yt-project.org/}} \citep{yt} to analyze our simulations and perform most figures.

\appendix
\section{Sink particles}
\label{sec:sink_a}

\begin{figure}
    \centering
    \includegraphics[width=0.5\columnwidth]{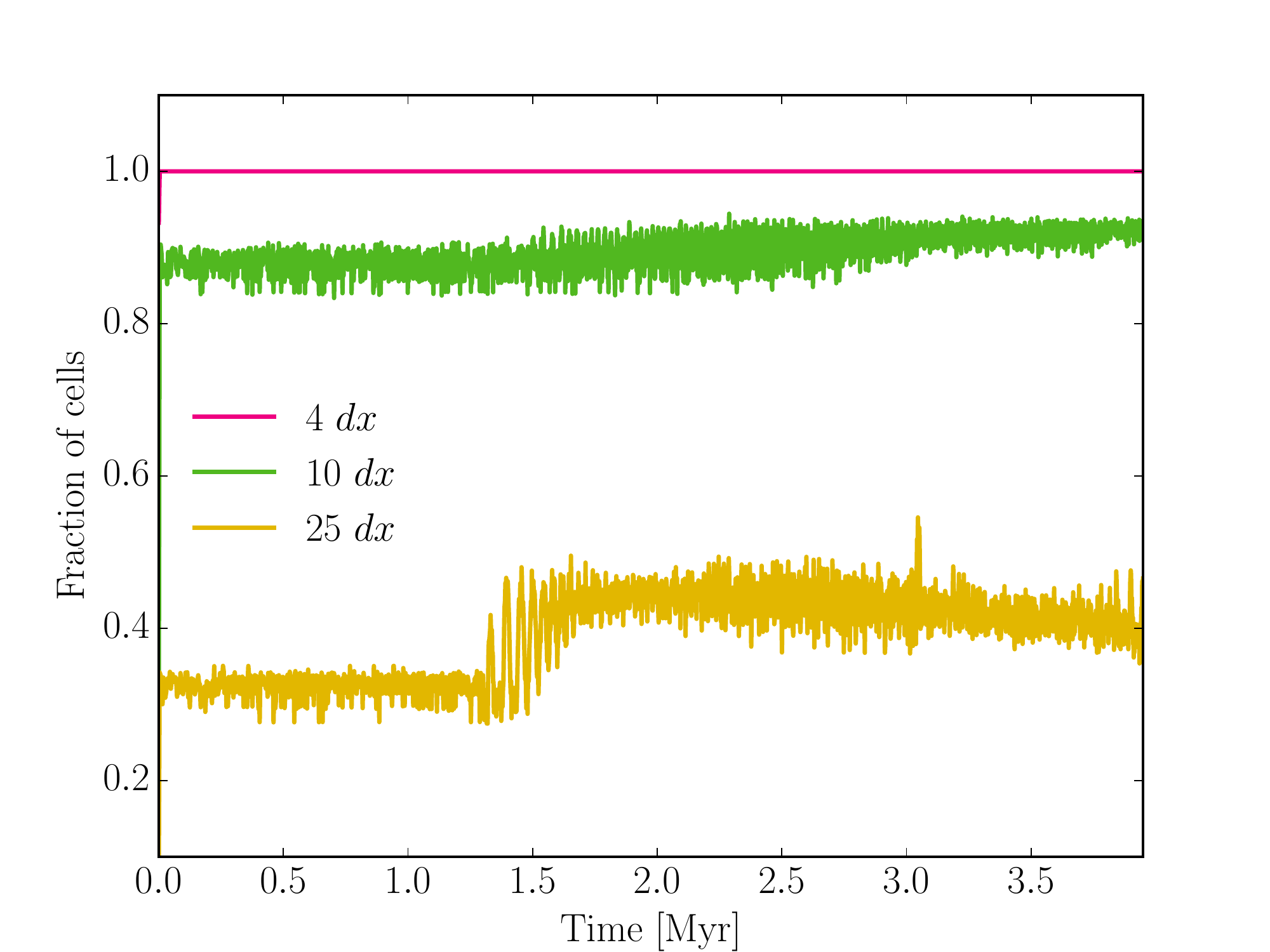}
    \caption{Fraction of cells at the maximum level of refinement as a function of time at different radii for the first sink particle that forms in Halo B at the $J_{21}$ = 10 run. The pink line represents the fraction considering a sphere of a radius of 4$dx$, green represents 10$dx$, and yellow represents it for 25$dx$, where $dx$ is the scale of the cells at the maximum level of refinement.}
    \label{fig:resol}
\end{figure}

In Section~\ref{sec:sink} we explain how our sink particles work, which is highly based on the work presented in \citet{bleuler}. In order to trust in the dynamics of our scheme, here we show some checks. \citet{Lupi} propose a new refinement criterion to follow in detail the motion of massive sink particles in AMR simulations with RAMSES. The criterion consists of enforcing to refine the region surrounding the sink particle. This refinement is motivated in physically following the trajectory of massive black hole binaries in the context of galaxy mergers. In Figure~\ref{fig:resol}, we show an example of the fraction of cells at the maximum level of refinement at three different radius of the sink particle that forms first in our halo B for the $J_{21}$ = 10 run. Pink represents a sphere of a radius of 4$dx$, green represents 10$dx$, while yellow represents 25$dx$, where $dx$ is the scale of the cell at the maximum level of refinement ($\sim$ 10$^{-2}$ pc). The behavior of all sink particles and their environments is so similar that this example applies to all the ones form in all our simulations. As we see, the fraction of cells at the maximum refinement at $4dx$ is one at all times, which is expected since $4dx$ is the accretion radius. The fraction of cells at $10dx$ is close to $0.9$, which is still considerable. Finally, at $25dx$ we see that the fraction is close to 0.3. It turns out that 25$dx$ is far enough to not have all cells at the maximum refinement level, but we see how it varies, as a result of the interaction of this particle with others, specially in binaries. From this check, we conclude that the amount of cells surrounding the sink particles at the maximum level of refinement satisfy the aforementioned criterion, so we can rely on their trajectories.

\begin{figure}
    \centering
    \includegraphics[width=0.9\columnwidth]{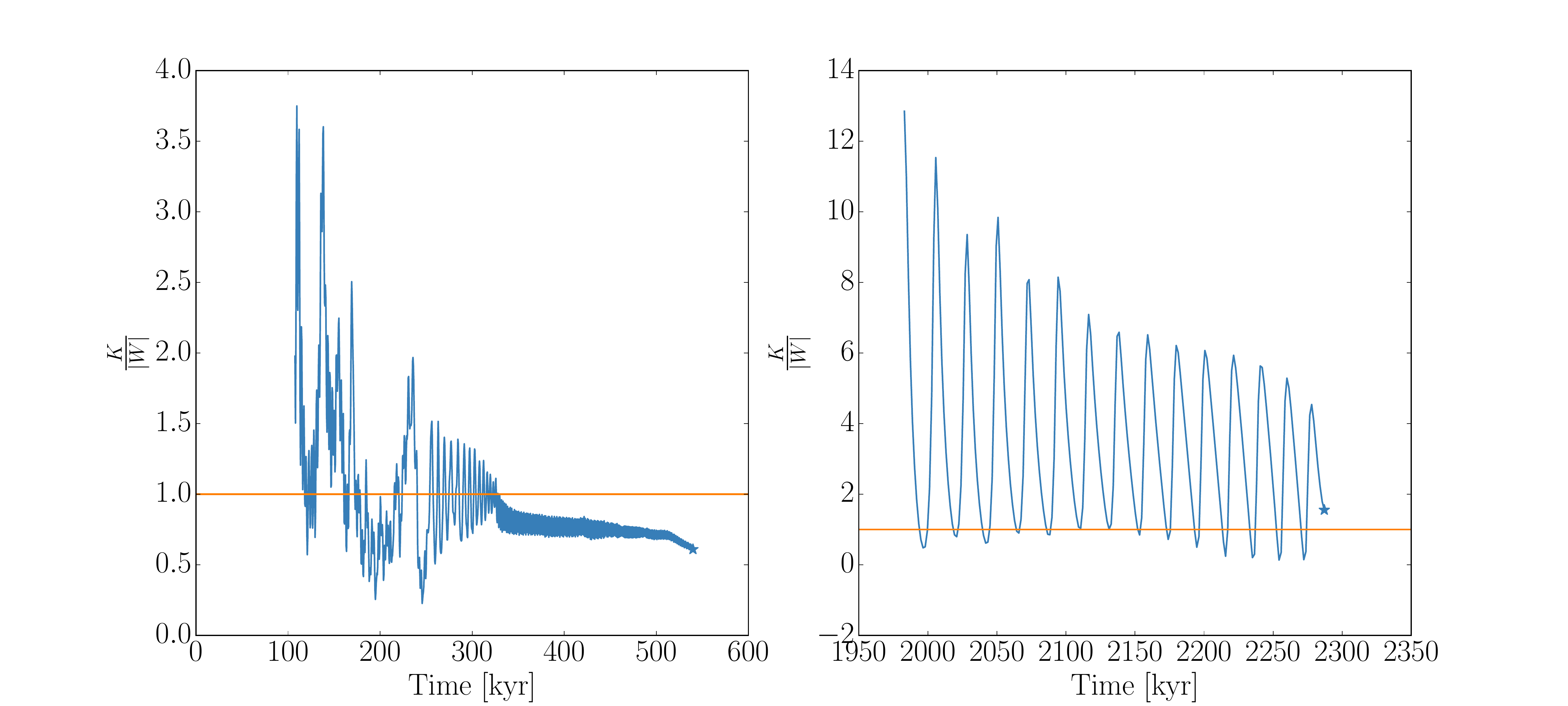}
    \caption{Kinetic energy over the absolute value of potential energy as a function of time for two pair of sink particles that merge. The left panel corresponds to the pair that merges in our Halo C under the presence of a UV background of $J_{21} = 10000$, while the right corresponds to Halo B under the presence of a UV intensity of $J_{21}$ = 10. In both cases, the merge is represented with a star and the orange line represents unity, which means that below that value the system is bound.}
    \label{fig:bound}
\end{figure}

In Section~\ref{sec:sink} we mention that sink particles merge if they are at a distance closer than four times the smallest scale. We study the energy of the pair of sink particles that merge, to know if they are bound at the moment they merge. In Figure~\ref{fig:bound} we show the quotient between the kinetic $K$ and the absolute value of the potential energy $|W|$ of two different systems of sink particles that merge. We choose this quantity since the bound condition $E < 0$, where $E$ is the total energy, is equivalent to the condition $K/|W| < 1$ when we neglect other sources of energy. In the left panel we show the energy of the system that merges in our Halo C run under the presence of a UV intensity of $J_{21} = 10000$. As we can see, the system is bound at the moment the merge happen. On the other hand, in the right panel we show the energy of the first system that merges in our halo B under the presence of a UV intensity of $J_{21} = 10$. In this case, the system becomes tighter as time goes on, but at the moment the particles meet, the condition in the energy for the system to be bound is not satisfied. The main reason is due to the fact that the energy has been calculated for the two sink particles in the system and we have not considered the amount of gas in the accretion radius at the moment they merge. We also have not considered that the merger happens in a triple system for this case. It seems that in our cases, the systems are bound at the moment they merge, even when the only condition to do it is numerical.

\section{Extended UV intensity cases}
\label{sec:j21}

In addition to the runs shown in Section~\ref{sec:results}, we also perform simulations with UV intensities of $J_{21} = 100$ and $J_{21} = 5$. There are no significant differences compared to the cases studied previously, they are mainly presented for completeness. 

\begin{figure}[ht]
\centering
\includegraphics[width=\columnwidth]{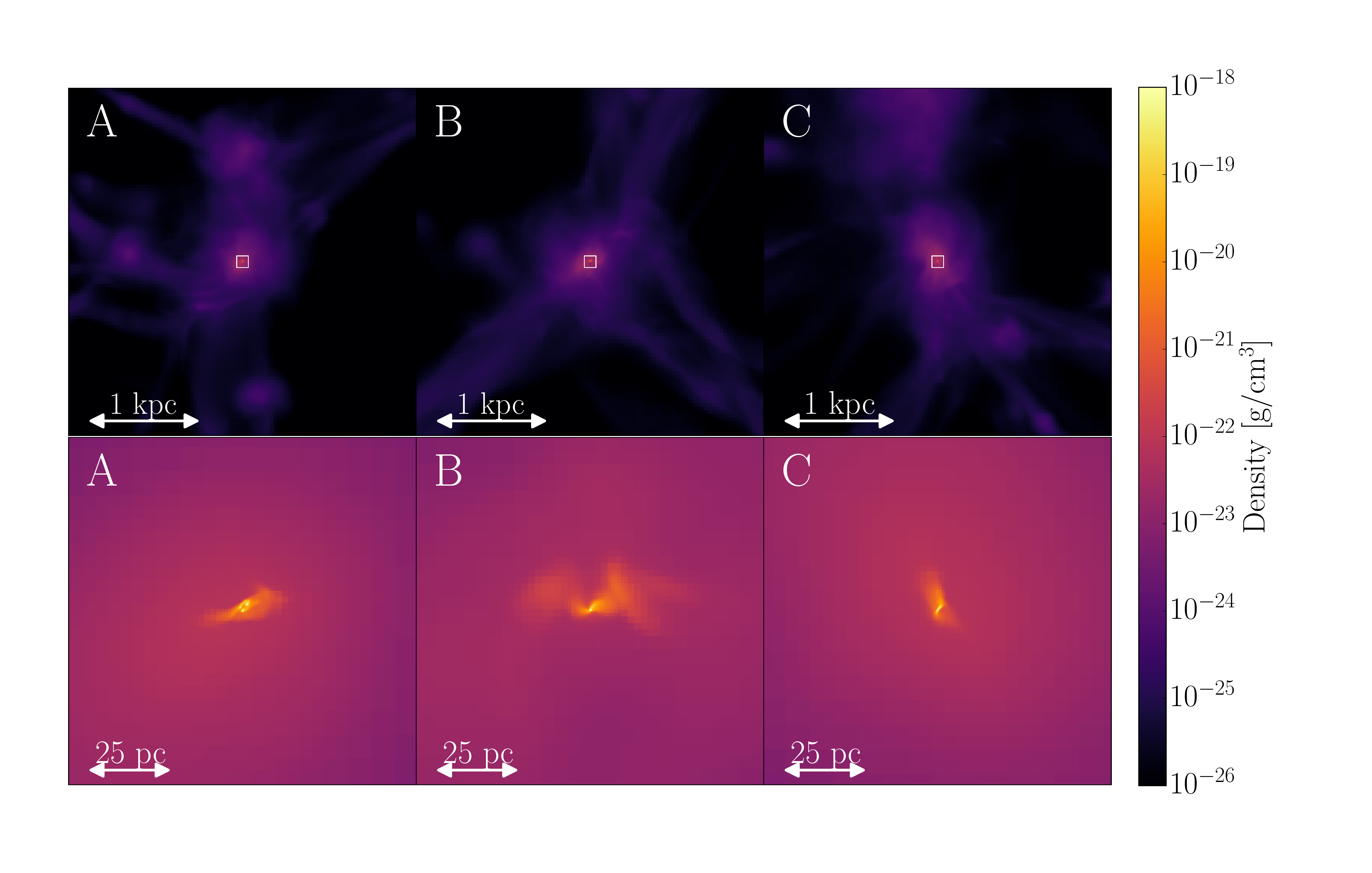}
\caption{Density projection for re-simulated Halos A, B, and C irradiated by a UV background of $J_{21} = 100$. The projections are made at the time the first sink particle is formed in the respective simulation. 
Halos A, B and C are shown from left to right, respectively. At the top, projections correspond to 3 kpc, where white squares correspond to 100 pc regions, which are shown at the bottom.}
\label{fig:j21_100}
\end{figure}

In Figure~\ref{fig:j21_100}, we can see density projections for Halos A, B, and C at scales of 3 kpc for the top panels and at scale of 100 pc for the bottom ones in the re-simulations performed under the presence of a intermediate UV background ($J_{21}$ = 100). In all cases just one sink particle forms, being simulated for $\sim 1$ Myr. In these runs, the sink particle masses range between the ones obtained in the $J_{21} = 10000$ and $J_{21} = 10$ cases. Additionally, gas distribution remains mainly central. Though all halos show some asymmetries, fragmentation and the formation of more than one sink particle is not observed during the period run. The formation of the only sink particle happens at a redshift of 12.06, 10.80, 11.77 for Halos A, B, and C, respectively. These values are in between the redshift where the collapse is triggered for $J_{21} = 10000$ and $J_{21} = 10$. These intermediate results in mass and redshift show that the formation and evolution of sink particles is related with gas thermal evolution, therefore, UV radiation has an impact on it.

The most relevant difference is that Halo C does not fragment nor produce more than one sink particle during the whole run, which seems inconsistent with the fragmentation observed under the presence of a stronger UV flux (Section~\ref{sec:j21_10000}). However, this inconsistency makes sense when we look at the redshifts at which the sink particles form and the merger trees (Figure~\ref{fig:merger_tree}). At $z$ = 10.80, the halo has suffered of a merger episode with another halo with a similar amount of mass, while at $z$ = 11.77, the halo was close to that merger episode, but it has not happened yet. This phenomenon again reveals the importance of dynamics and the environment in which interactions are happening.

In addition to the re-simulations under the presence of a UV flux with a $J_{21}$ = 100, We also performed one re-simulation for Halo B under the influence of a $J_{21} = 5$ UV background. We picked Halo B, since various interactions between the sink particles formed are observed in its $J_{21} = 10$ run. This run includes deuteride species and their associated reactions, which are shown in Appendix~\ref{sec:chem}.  

In Figure~\ref{fig:low_5} we can see different density projections of Halo B in the setup mentioned above, the structure in this run is too similar to the re-simulation of Halo B, the filamentary structure is distributed similarly, nine sink particles are created, and two merger episodes are observed. All surviving sink particles end up with masses between 10$^{3}$ and 10$^{4}$ $M_{\odot}$.

\begin{figure}
    \centering
    \includegraphics[width=0.65\columnwidth]{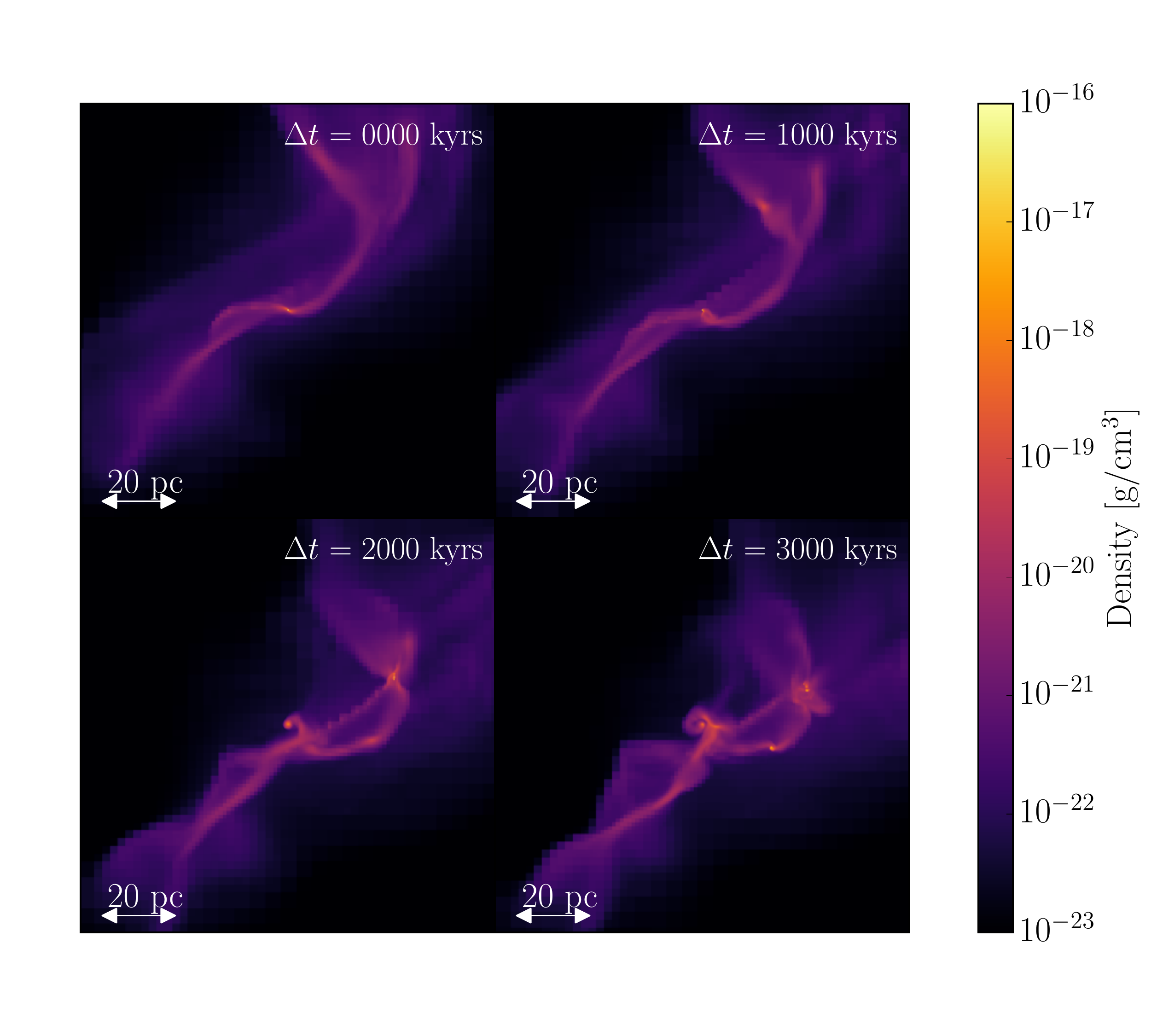}
    \caption{Density projections for the re-simulation of Halo B under the influence of a uniform UV background of $J_{21} = 5$ that includes a deuterium chemistry.}
    \label{fig:low_5}
\end{figure}

\section{Accretion Rates}
\label{sec:accretion}

In Section~\ref{sec:sink_j21_10}, we revise accretion rates for Halos A and B. In Figure~\ref{fig:sink_10_v_1000}, we apply a smoothing technique to better visualize the differences between sink particles and to focus on the long-term variations. By doing this, we lose relevant information about the oscillations in accretion rates due to gravitational interactions between sink particles. For instance, \citet{goicovic} show that for black hole binaries, accretion rates oscillate with the rotation timescale of the system.
 
\begin{figure}
\centering
\includegraphics[width=0.9\columnwidth]{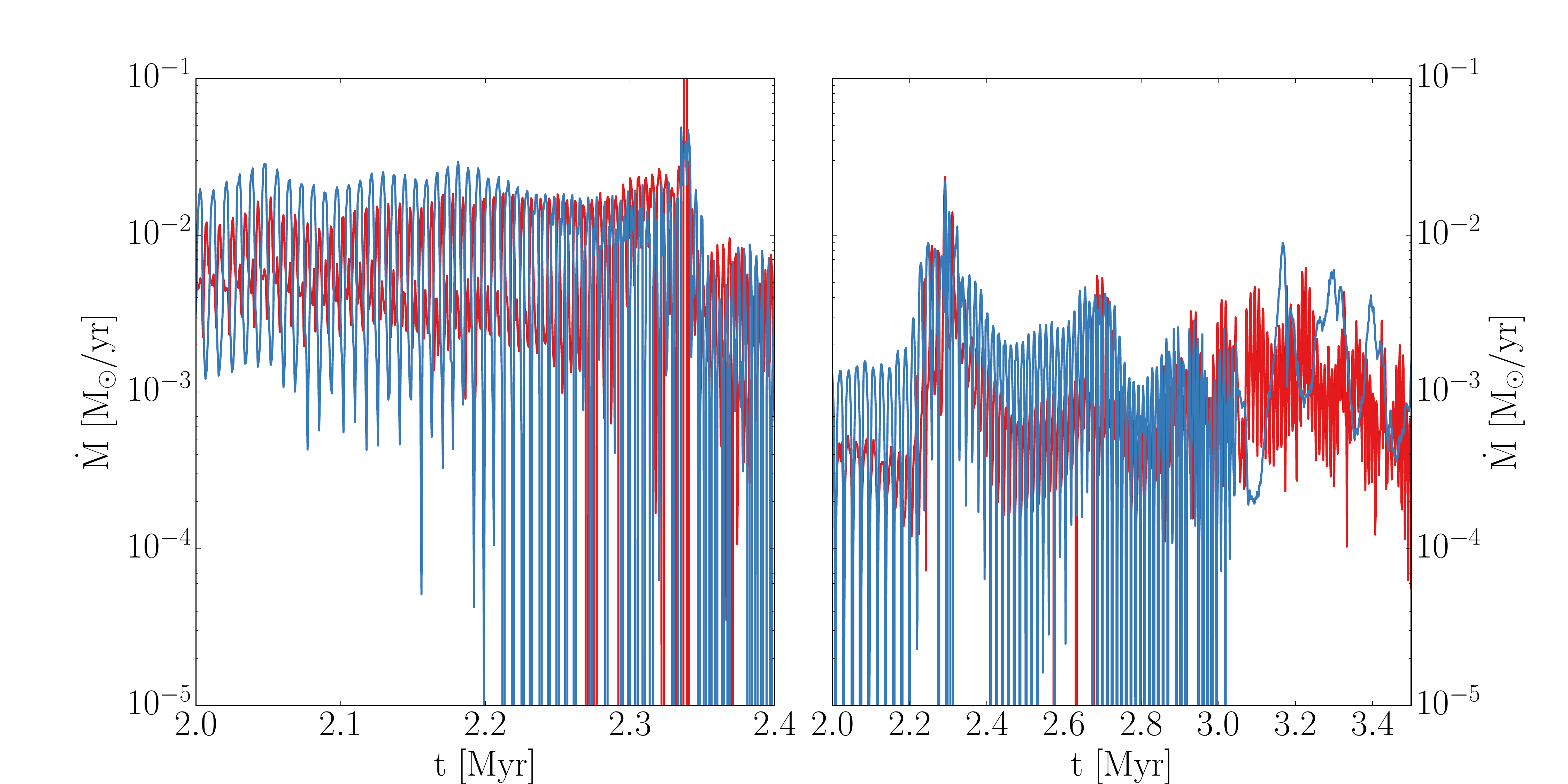}
\caption{Accretion rate time evolution with no smoothing technique in the left panel for Halo A and in the right panel for Halo B. Every color represents a sink particle, they are the same used in Figure~\ref{fig:sink_10_v_1000} for Halo A, and in Figure~\ref{fig:sink_10_v_1000} for Halo B. In both cases, the accretion rates correspond to the runs with $J_{21}$ = 10.}
\label{fig:acc_10}
\end{figure}

In Figure~\ref{fig:acc_10}, we show accretion rates evolution with time for Halo A in the left panel and for Halo B in the right one; in both cases the times in the $x$-axis are relative to the first sink particle formed in each respective run. For Halo A, this evolution ranges between $\Delta t =$ 2.0 and 2.4 Myr, while for B it ranges between $\Delta t =$ 2.0 and 3.4 Myr. No smoothing technique has been applied, so we appreciate raw values. For Halo A, we observe an oscillatory behavior that is not periodic, since they become more frequent with time due to the particles approaching. We also observe extreme variations, since the values oscillate between values $< 10^{-5}$ $M_{\odot}$yr$^{-1}$ and $> 10^{-2}$ $M_{\odot}$yr$^{-1}$, especially for the least massive member of the binary (blue). For Halo B, we also observe an oscillatory behavior. Unlike Halo A, we see three `phases', one from $\Delta t \leq 2.0$ to $2.2$ Myr, another from $\Delta t = 2.2$ to $3.0$ Myr, and a third one from $\Delta t = 3.0$ to $> 3.5$ Myr. The first phase is related to the binary interaction between both members in the group. The behavior is similar to Halo A, since there are only two particles interacting. The second `phase' is related to the approach of a dense region that forms a sink particle (purple line in Figure~\ref{fig:sink_10_v_1000}), which slightly enhances the accretion rate. It is interesting to note that the first change in the accretion rate by $\Delta t$ = 2.2 Myr begins before a sink particle is created, which merges with one of the binary members, later, the same behavior is appreciated at $\Delta t$ = 2.6 Myr once a new dense core approaches to the binary system. Finally, we find a third regime, which starts after the formation of the triple system between the already formed binary system and the sink that forms in the core that previously have raised the accretion. The accretion rate of the most massive member in the old binary does not change considerably, while the least massive behaves more smoothly compared and its previous history and keeps its accretion rates in a more constrained range.

\section{Chemical Reactions}
\label{sec:chem}

All of the chemical reactions used and their respective rate coefficients  are listed in Table~\ref{tab:long} along with references. In rate coefficients, $T$ is gas temperature in K, $T_e$ is gas temperature in eV, $T_{rad}$ is radiation temperature in K, $J_{21}$ is the UV intensity in units of $J_{21}$ (10$^{-21}$ erg s$^{-1}$ Hz$^{-1}$ cm$^{-2}$ str$^{-1}$), and $n_{H}$ is number density for neutral hydrogen in n cm$^{-3}$.

Chemical reactions (1)-(28) are found in the \textit{react\_xrays} network included in the KROME package. This network includes the most predominant reactions in a primordial environment, also existing literature about the rates to create/destroy them. We also include reactions (29) and (30) from the minimal model derived by \citet{glover2015}.

Deuterium and molecules containing this element are expected to be outcomes of the nucleosynthesis; however, their abundances are so low that the cooling from these species is irrelevant compared to molecular hydrogen cooling. They are potentially important in environments with sufficient electron fraction to catalyze the HD formation \citep{shchekinov}. The simulations performed in the main body of this paper do not include these chemical species due to the aforementioned reasons and to save computational memory. We perform one simulation including a deuterium chemistry to test its relevance (Appendix~\ref{sec:j21}) and as a check to the dissociation assumption. The following species were added: D, D−, HD, D+, HD+. The chemical reactions and their coefficient rates and references are shown in Table~\ref{tab:deuteride}.


\newpage

\begin{center}
\begin{longtable}{p{4cm}p{11cm}p{1cm}}
\caption{Chemical reactions with their respective rate coefficient and references. Rate coefficients are in units of cm$^{3(n-1)}$s$^{-1}$ were $n$ is the number of reactants. Photo-dissociation's rate coefficients are in units of 1/s.} \label{tab:long} \\

\hline \multicolumn{1}{c}{\textbf{Reaction}} & \multicolumn{1}{c}{\textbf{Rate Coefficient}} & \multicolumn{1}{c}{\textbf{Ref.}} \\ \hline 
\endfirsthead

\hline \multicolumn{1}{c}{\textbf{Reaction}} & \multicolumn{1}{c}{\textbf{Rate coefficient}} & \multicolumn{1}{c}{\textbf{Ref.}} \\ \hline 
\hline
\endhead


\endlastfoot
(1) H + e$^{-}$ $\rightarrow$ H$^{+}$ + 2e$^{-}$ & $k_1 = \exp[-32.71396786$ & \multicolumn{1}{c}{1}\\
& $+13.5365560$ $\ln T_e$  & \\
& $-5.73932875$ $(\ln T_e)^2$ & \\
& $+1.56315498$ $(\ln T_e)^3$ & \\
& $-0.28770560$ $(\ln T_e)^4$ & \\
& $+3.48255977 \times 10^{-2}$ $(\ln T_e)^5$ & \\
& $-2.63197617 \times 10^{-3}$ $(\ln T_e)^6$ & \\ 
& $+1.11954395 \times 10^{-4}$ $(\ln T_e)^7$ & \\
& $-2.03914985 \times 10^{-6}$ $(\ln T_e)^8]$ & \\
(2) H$^{+}$ + e$^{-}$ $\rightarrow$ H + $\gamma$& $k_2 = 3.92 \times 10^{-13}$ $T_e^{-0.6353}$ \hfill $T \leq 5000$ K & \multicolumn{1}{c}{2}\\
&$k_2 = \exp[-28.61303380689232$ \hfill $T > 5500$ K& \\
&$-7.241125657826851\times 10^{-1}$ $\ln T_e$ & \\
&$-2.026044731984691\times 10^{-2}$ $(\ln T_e)^2$ & \\
&$-2.380861877349834\times 10^{-3}$ $(\ln T_e)^3$ & \\
&$-3.212605213188796\times 10^{-4}$ $(\ln T_e)^4$& \\
&$-1.421502914054107\times 10^{-5}$ $(\ln T_e)^5$& \\
&$+4.989108920299513\times 10^{-6}$ $(\ln T_e)^6$& \\
&$+5.755614137575758\times 10^{-7}$ $(\ln T_e)^7$& \\
&$-1.856767039775261\times 10^{-8}$ $(\ln T_e)^8$& \\
&$-3.071135243196595\times 10^{-9}$ $(\ln T_e)^9]$& \\
(3) He + e$^{-}$ $\rightarrow$ He$^+$ + 2e$^-$ & $k_3 =\exp[-44.09864886$ \hfill & \multicolumn{1}{c}{1} \\ 
&$+23.91596563$ $\ln T_e$ & \\
&$-10.7532302$ $(\ln T_e)^2$ & \\
&$+3.05803875$ $(\ln T_e)^3$ & \\
&$-0.56851189$ $(\ln T_e)^4$ & \\
&$+6.79539123 \times 10^{-2}$ $(\ln T_e)^5$ & \\
&$-5.00905610 \times 10^{-3}$ $(\ln T_e)^6$ & \\
&$+2.06723616 \times 10^{-4}$ $(\ln T_e)^7$ & \\
&$-3.64916141 \times 10^{-6}$ $(\ln T_e)^8]$ & \\
(4) He$^+$ + e$^-$ $\rightarrow$ He + $\gamma$ & $k_4 = 3.92\times 10^{-13}$ $T_e^{-0.6353}$ \hfill $T \leq 9280$ K & \multicolumn{1}{c}{3} \\
& $k_4 = 3.92\times 10^{-13}$ $T_e^{-0.6353}$ \hfill $T > 9280$ K & \\
& $+$ $1.54\times 10^{-9}$ $T_e^{-1.5}[1+0.3/\exp(8.099328789667/T_e)]$ &  \\
& $/[\exp(40.49664394833662/T_e)]$& \\
(5) He$^+$ + e$^-$ $\rightarrow$ He$^{++}$ + 2e$^-$ & $k_5 = \exp[-68.71040990212001$  & \multicolumn{1}{c}{4}\\
&$+43.93347632635$ $\ln T_e$&\\
&$-18.48066993568$ $(\ln T_e)^2$&\\
&$+4.701626486759002$ $(\ln T_e)^3$&\\
&$-7.692466334492 \times 10^{-1}$ $(\ln T_e)^4$&\\
&$+8.113042097303 \times 10^{-2}$ $(\ln T_e)^5$&\\
&$-5.324020628287001 \times 10^{-3}$ $(\ln T_e)^6$&\\
&$+1.975705312221 \times 10^{-4}$ $(\ln T_e)^7$&\\
&$-3.165581065665 \times 10^{-6}$ $(\ln T_e)^8)]$& \\
\hline
\multicolumn{3}{c}{\vspace{-2.2mm}}\\
\multicolumn{3}{c}{1: \citet{janev}, 2: \citet{abel} fit by data from \citet{ferland}}\\
\multicolumn{3}{c}{3: \citet{cen}; \citet{aldrovandi}, 4: Aladdin database, see \citet{abel}}\\

\newpage

(6) He$^{++}$ + e$^-$ $\rightarrow$ He$^+$ + $\gamma$ & $k_6 = 1.891\times 10^{-10}/a$ & \multicolumn{1}{c}{5} \\
& $a = (1.0+\sqrt{T/9.37})^{0.2476}(1.0+\sqrt{T/2.774 \times 10^6})^{1.7524}\sqrt{T/9.37}$ & \\
(7) H$^+$ + e$^-$ $\rightarrow$ H$^-$ + $\gamma$ & $k_7 = 1.4\times 10^{-18}$ $T^{0.928} \exp(-T/16200)$ & \multicolumn{1}{c}{6} \\
(8) H$^-$ + H $\rightarrow$ H$_2$ + e$^-$& $k_8 = a_1(T^{a_2}+a_3T^{a_4}+a_5T^{a_6})/(1.0+a_7T^{a_8}+a_9T^{a_{10}}+a_{11}T^{a_{12}})$ & \multicolumn{1}{c}{7}\\
&$a_1=1.3500\times 10^{-9}$&\\
&$a_2=9.8493\times 10^{-2}$&\\
&$a_3=3.2852\times 10^{-1}$&\\
&$a_4=5.5610\times 10^{-1}$&\\
&$a_5=2.7710\times 10^{-7}$&\\
&$a_6=2.1826$&\\
&$a_7=6.1910\times 10^{-3}$&\\
&$a_8=1.0461$&\\
&$a_9=8.9712\times 10^{-11}$&\\
&$a_{10}=3.0424$&\\
&$a_{11}=3.2576\times 10^{-14}$&\\
&$a_{12}=3.7741$&\\
(9) H + H$^+$ $\rightarrow$ H$_2^{+}$ + $\gamma$& $k_9 = 2.10 \times 10^{-20}$ $(T/30)^{-0.15}$ \hfill $T < 30$ K & \multicolumn{1}{c}{8} \\
& $k_9 = \text{dex}[-18.20-3.194 \log_{10}T+ $ \hfill $T > 30$ K &\\
 & $1.786(\log_{10}T)^2-0.2072(\log_{10}T)^3]$ & \\
(10) H$_2^{+}$ + H $\rightarrow$ H$_2$ + H$^+$ & $k_{10} = 6.0 \times 10^{-10}$ & \multicolumn{1}{c}{9} \\
(11) H$_2$ + H$^+$ $\rightarrow$ H$_2^+$ + H & $k_{11} = \exp(-a/T)[b_1+b_2\ln T+b_3 (\ln T)^2$ \hfill $100$ K $\leq T \leq 30000$ K & \multicolumn{1}{c}{10}\\ 
& $+b_4 (\ln T)^3+b_5(\ln T)^4 +b_6(\ln T)^5 +b_7(\ln T)^6 + b_8(\ln T)^7 ]$  & \\
&$a = 2.1237150 \times 10^{4}$&\\
&$b_1=-3.3232183\times 10^{-7}$&\\
&$b_2= 3.3735382\times 10^{-7}$&\\
&$b_3=-1.4491368\times 10^{-7}$&\\
&$b_4= 3.4172805\times 10^{-8}$&\\
&$b_5=-4.7813728\times 10^{-9}$&\\
&$b_6= 3.9731542\times 10^{-10}$&\\
&$b_7=-1.8171411\times 10^{-11}$&\\
&$b_8= 3.5311932\times 10^{-13}$&\\
(12) H$_2$ + e$^-$ $\rightarrow$ H + H$^-$ & $k_{12} = 35.5$ $T^{-2.28}$ $\exp(-46707/T)$ & \multicolumn{1}{c}{11}\\
(13) H$_2$ + e$^-$ $\rightarrow$ H + H + e$^-$& $k_{13} = 4.38 \times 10^{-10}$ $T^{0.35}$ $\exp(-102000/T)$  & \multicolumn{1}{c}{12}\\
(14) H$_2$ + H $\rightarrow$ H + H + H & $k_{14}$, see ref. & \multicolumn{1}{c}{13}\\
(15) H$^-$ + e$^-$ $\rightarrow$ H + 2e$^-$ & $k_{15} = \exp[-18.01849334273$ & \multicolumn{1}{c}{1} \\
&$+2.360852208681$ $\ln T_e$&\\
&$-2.827443061704 \times 10^{-1} $ $(\ln T_e)^2$&\\
&$+1.623316639567 \times 10^{-2}$ $(\ln T_e)^3$&\\
&$-3.365012031362999 \times 10^{-2}$ $(\ln T_e)^4$&\\
&$+1.178329782711 \times 10^{-2}$ $(\ln T_e)^5$&\\
&$-1.656194699504 \times 10^{-3}$ $(\ln T_e)^6$&\\
&$+1.068275202678 \times 10^{-4}$ $(\ln T_e)^7$&\\
&$-2.63128580920710 \times 10^{-6}$ $(\ln T_e)^8]$ & \\

\hline
\multicolumn{3}{c}{\vspace{-2.2mm}}\\
\multicolumn{3}{c}{5: \citet{verner}, 6: \citet{dejong}, 7: \citet{kreckel}}\\
\multicolumn{3}{c}{8: \citet{coppola}, 9: \citet{karpas}}\\
\multicolumn{3}{c}{10: \citet{savin}, 11: \citet{capitelli}}\\
\multicolumn{3}{c}{12: \citet{mitchell}, fit of data by \citet{corrigan}, 13: \citet{martin}}\\

\newpage

(16) H$^-$ + H $\rightarrow$ H + H + e$^-$& $k_{16} = 2.5610 \times 10^{-9}$ $T_e^{1.781860}$ \hfill $T \leq 1160$ K & \multicolumn{1}{c}{14} \\
&$k_{16} = \exp[-20.37260896533324$ \hfill $T > 1160$ K&\\
&$+1.139449335841631 \ln T_e$ &\\
&$-1.421013521554148 \times 10^{-1}$ $(\ln T_e)^2$ &\\
&$+8.46445538663 \times 10^{-3}$ $(\ln T_e)^3$ &\\
&$-1.4327641212992 \times 10^{-3}$ $(\ln T_e)^4$ &\\
&$+2.012250284791 \times 10^{-4}$ $(\ln T_e)^5$ &\\
&$+8.66396324309 \times 10^{-5}$ $(\ln T_e)^6$ &\\
&$-2.585009680264 \times 10^{-5}$ $(\ln T_e)^7$ &\\
&$+2.4555011970392 \times 10^{-6}$ $(\ln T_e)^8$ &\\
&$-8.06838246118 \times 10^{-8}$ $(\ln T_e)9$ &\\
(17) H$^-$ + H$^+$ $\rightarrow$ H + H & $k_{17} = 2.96 \times 10^{-6}/\sqrt{T}-1.73 \times 10^{-9}$ \hfill $10$ K $< T < 10^{5}$ K  & \multicolumn{1}{c}{15}\\
& $+ 2.5 \times 10^{-10} \sqrt{T} - 7.77 \times 10^{-13}T$ & \\
(18) H$^-$ + H$^+$ $\rightarrow$ H$_2^+$ + e$^-$ & $k_{18} = 10^{-8}$ $T^{-0.4}$ & \multicolumn{1}{c}{16} \\
(19) H$^-$ + $\gamma$ $\rightarrow$ H + e$^-$& $k_{19} = J_{21} \times$ $10^{[(a + bT_{\text{rad}})^{-1/c} - d]}
$ & \multicolumn{1}{c}{17} \\
&$a = 9.08944 \times 10^{-2}$&\\
&$b = 3.27940 \times 10^{-5}$&\\
&$c = 5.98490 \times 10^{-1}$&\\
&$d = 10.9867$&\\(20) H$_2$ + $\gamma$ $\rightarrow$ H + H & $k_{20} = f_{sh}(n,T) \times J_{21} \times 10^{[1.0 / (a + bT_{\text{rad}} + cT_{\text{rad}} ^2 ) - d]}$ & \multicolumn{1}{c}{18}\\
&$a = 1.17350 \times 10^{-1}$&\\
&$b = 2.49580 \times 10^{-4}$&\\
&$c = 3.48559 \times 10^{-9}$&\\
&$d = 11.902$&\\
(21) H$_2^+$ + $\gamma$ $\rightarrow$ H$^+$ + H & $k_{21} = J_{21} \times [(a + bT_{\text{rad}})^{-1/c} + d]$
 & \multicolumn{1}{c}{18} \\
&$a = -3.83012 \times 10^{6}$&\\
&$b = 5.06447 \times 10^{2}$&\\
&$c = 6.20988 \times 10^{-1}$&\\
&$d = 3.68778 \times 10^{-12}$&\\
(22) H$_2^{+}$ + e$^-$ $\rightarrow$ H + H + $\gamma$
&$k_{22} = 10^{6}[4.2278 \times 10^{-14}$ \hfill $T \leq 10000$ K& \multicolumn{1}{c}{19}\\
&$-2.3088 \times 10^{-17}T$&\\
&$+7.3428 \times 10^{-21}T^2$&\\
&$-7.5474 \times 10^{-25}T^3$&\\
&$+3.3468 \times 10^{-29}T^4$&\\
&$-5.528 \times 10^{-34}T^5]$&\\
(23) H$_2^{+}$ + H$^-$ $\rightarrow$ H + H$_2$ & $k_{23} = 5 \times 10^{-7}$ $\sqrt{100/T}$
 & \multicolumn{1}{c}{20}\\
(24) 3H $\rightarrow$ H$_2$ + H & $k_{24} = 6\times10^{-32}T^{-0.25}+2\times 10^{-31}$ $T^{-0.5}$
 & \multicolumn{1}{c}{21}\\
(25) H$_2$ + 2H $\rightarrow$ 2H$_2$ & $k_{25} = k_{24}/8$ & \multicolumn{1}{c}{22}\\
(26) 2H$_2$ $\rightarrow$ 2H + H$_2$ & $k_{26} = kh_{26}^{(1.0-a_{26})} \times kl_{26}^{a_{26}}$
 & \multicolumn{1}{c}{23}\\
&$kl_{26} = 1.18 \times 10^{-10} \exp(-6.95 \times 10^{4}/T)$&\\
&$kh_{26} = 8.125 \times 10^{-8}$ $T^{-0.5}$ $\exp(-5.2 \times 10^{4}/T)$ $(1.0-\exp (-6\times 10^{3}/T))$&\\
&$ncr_{26} = \text{dex}[4.845-1.3 \log (T/10^{4})+1.62[\log (T/10^{4})]^2]$&\\
&$a_{26} = 1/(1+(\text{nH}/ncr_{21}))$&\\

\hline
\multicolumn{3}{c}{\vspace{-2.2mm}}\\
 \multicolumn{3}{c}{14: \citet{abel} based on \citet{janev}, 15: \citet{stenrup}}\\
 \multicolumn{3}{c}{16: \citet{poulaert}, 17: \citet{latif2015c}, see also \citet{miyake}}\\
 \multicolumn{3}{c}{18: \citet{latif2015c}, 19: \citet{coppola}, 20: \citet{dalgarno}}\\
\multicolumn{3}{c}{21: \citet{forrey}, 22: adopted from \citet{forrey},23: \citet{omukai2001}}\\

\newpage

(27) He$^+$ + H $\rightarrow$ He + H$^+$ & $k_{27} = 1.2\times 10^{-15}$ $(T/300)^{0.25}$ & \multicolumn{1}{c}{24}\\
(28) He + H$^+$ $\rightarrow$ He$^+$ + H & $k_{28} = 1.26 \times10^{-9}$ $T^{-0.75}\exp(-1.275 \times 10^{5}/T)$ \hfill $T \leq 10^4$ K & \multicolumn{1}{c}{24}\\
&$k_{28} = 4 \times 10^{-37} T^{4.74} $ \hfill $T > 10^4$ K &\\
(29) 2H $\rightarrow$ H$^+$ + e$^-$ + H & $k_{29} = 1.2 \times 10^{-17}$ $T^{1.2}\exp(-1.578 \times 10^{5}/T)$ & \multicolumn{1}{c}{25} \\
(30) H + He $\rightarrow$ H$^+$ + e$^-$ + He & $k_{30} = 1.75 \times 10^{-17}T^{1.3}\exp(-1.578\times 10^{5}/T)$
 & \multicolumn{1}{c}{25}\\
\hline
\multicolumn{3}{c}{\vspace{-2.2mm}}\\
\multicolumn{3}{c}{24: \citet{yoshida2006}, 25: \citet{glover2015}}\\
\end{longtable}
\end{center}

\begin{center}
\begin{longtable}{p{4cm}p{11cm}p{1cm}}
\caption{Chemical reactions for deuteride species with their respective rate coefficients and references. Rate coefficients are in units of cm$^3$/s.} \label{tab:deuteride} \\

\hline \multicolumn{1}{c}{\textbf{Reaction}} & \multicolumn{1}{c}{\textbf{Rate Coefficient (cm$^3$s$^{-1}$)}} & \multicolumn{1}{c}{\textbf{Ref.}} \\ \hline 
\endfirsthead

\hline \multicolumn{1}{c}{\textbf{Reaction}} & \multicolumn{1}{c}{\textbf{Rate coefficient}} & \multicolumn{1}{c}{\textbf{Ref.}} \\ \hline 
\hline
\endhead


\endlastfoot
(31) H$^{+}$ + D $\rightarrow$ H + D$^{+}$ & $k_{31} = 2.00 \times 10^{-10}T^{0.402}\exp(-37.1/T) - 3.31 \times 10^{-17} T^{1.48}$ \hfill T $\geq 50 $ K & 26\\
(32) H + D$^{+}$ $\rightarrow$ H$^{+}$ + D & $k_{32} = 2.06 \times 10^{-10}T^{0.396} \exp(-33/T) + 2.03 \times 10^{-9}T^{-0.332}$ \hfill T $\geq 50$ K & 26\\
(33) H$_2$ + D$^+$ $\rightarrow$ HD + H$^+$ & $k_{33} = 10^{-9}[0.417+0.846\log _{10}(T)-0.137(\log _{10}(T))^2]$ & 27\\
(34) HD + H$^+$ $\rightarrow$ H$_2$ + D$^+$ & $k_{34} = \exp(-457/T) \times 10^{-9} $ & 28\\
(35) H$_2$ + D $\rightarrow$ HD + H & $k_{35} = \text{dex}[-56.4737 + 5.88886 \log _{10}T $ \hfill T $\leq 2000$ K & 29\\
 & $+ 7.19692 (\log _{10}T)^2 + 2.25069( \log _{10}T)^3  $ & \\
 & $- 2.16903( \log _{10}T)^4 + 0.317887(\log _{10}T)^5] $ & \\
 & $k _{35} =  3.17 \times 10^{-10}\exp(-5207/T) $ \hfill T $> 2000$ K& \\
(36) HD + H $\rightarrow$ H$_2$ + D & $k _{36} = 5.25 \times 10^{-11}\exp(-4430/T+1.739\times 10^5/T^2)$ \hfill T $> 200$ K & 30\\
(37) D + H$^{-}$ $\rightarrow$ HD + e$^{-}$ & $k_{37} = 1.5 \times 10^{-9}(T/300)^{-0.1}$ & 31\\
(38) D$^+$ + e$^{-}$ $\rightarrow$ D & $k_{38} = 3.6 \times 10^{-12}(T/300)^{-0.75}$ & 31\\
(39) H + D $\rightarrow$ HD & $k_{39} = 10^{-25} $ & 32\\
(40) HD$^+$ + H $\rightarrow$ HD + H$^+$ & $k_{40} = 6.4 \times 10^{-10} $ & 31,32\\
(41) H$^+$ + D $\rightarrow$ HD$^+$ & $k_{41} = \text{dex}[-19.38 - 1.523\log _{10}T $ & 32\\
& $+ 1.118(\log _{10}T)^2 -0.1269(\log _{10}T)^3]$ & \\
(42) HD$^+$ + e$^{-}$ $\rightarrow$ H + D & $k_{42} = 7.2\times 10^{-8}/\sqrt{T} $ & 32\\
(43) D + e$^{-}$ $\rightarrow$ D$^{-}$ & $k_{43} = 3\times 10^{-16} (T/300)^{0.95}\exp(-T/9320) $ & 31\\
(44) D$^+$ + D$^{-}$ $\rightarrow$ D + D & $k_{44} = 5.7 \times 10^{-8}(T/300)^{-0.5}$ & 31\\
(45) D$^-$ + H$^{+}$ $\rightarrow$ D + H & $k_{45} = 4.6 \times 10^{-8}(T/300)^{-0.5}$ & 31\\
(46) H$^-$ + D $\rightarrow$ D$^{-}$ + H & $k_{46} = 6.4 \times 10^{-9}(T/300)^{0.41}$ & 31\\
(47) H$^-$ + D $\rightarrow$ D$^{-}$ + H & $k_{47} = 6.4 \times 10^{-9}(T/300)^{0.41}$ & 31\\
(48) D$^-$ + H $\rightarrow$ HD + e$^-$ & $k_{48} = 1.5 \times 10^{-9}(T/300)^{-0.1}$ & 31\\
\hline
\multicolumn{3}{c}{\vspace{-2.2mm}}\\
 \multicolumn{3}{c}{26: \citet{galli_2002} from \citet{savin_2001}, 27: \citet{galli_2002} from \citet{gerlich}}\\
  \multicolumn{3}{c}{28: \citet{galli_2002} from \citet{gerlich} with modification by \citet{gay_2011}}\\
\multicolumn{3}{c}{29: \citet{glover_2009} from data by \citet{mielke_2003}, 30: \citet{galli_2002}, from \citet{shavitt_1959}}\\
\multicolumn{3}{c}{31: \citet{Stancil}, 32: \citet{galli_98}}\\
\end{longtable}
\end{center}

\bibliography{example.bib}

\end{document}